\documentclass[twocolumn]{aa}
\usepackage{graphicx}
\usepackage{txfonts}
\usepackage{natbib}
\include{definitions}

\bibpunct{(}{)}{;}{a}{}{,} % to follow the A&A style

\begin{document}
   \title{Radio and X-ray variability of Young Stellar Objects \\ in the
\textsl{\textbf{Coronet}} cluster}

%   \subtitle{}

\author{J. Forbrich\inst{1,2}, Th. Preibisch\inst{1}, \and K. M. Menten\inst{1}}

\offprints{J. Forbrich\\
\email{forbrich@mpifr-bonn.mpg.de}}

\institute{Max-Planck-Institut f\"ur Radioastronomie, Auf dem H\"ugel 69, D-53121 Bonn, Germany \and Astrophysikalisches Institut und Universit\"ats-Sternwarte Jena, Schillerg\"a{\ss}chen 2-3, D-07745 Jena, Germany}

   \date{Received / Accepted}

   \abstract{The \textsl{Coronet} Cluster in the nearby R~CrA dark
cloud offers the rare opportunity to study at least four ``class~I''
protostellar sources as well as one candidate ``class~0'' source, a Herbig Ae star, and a candidate brown dwarf within a few square arcminutes -- most of them detected at radio- \textsl{and} X-ray wavelengths. These sources were observed with the Very Large Array (VLA) at $\lambda = 3.5$~cm on nine occasions in 1998, spread over nearly four months. The source IRS~5, earlier shown to emit circularly polarized radio emission, was observed to undergo a flux increase accompanied by changes in its polarization properties. Comparison with VLA measurements taken in January 1997 allows for some analysis of longer-term variability. In addition to this radio monitoring, we analyze archival \textsl{Chandra} and XMM-\textsl{Newton} X-ray data of these sources. Three class~I protostars are bright enough for X-ray spectroscopy, and we perform a variability analysis for these sources, covering a total of 154~ksec spread over more than two and a half years. Also in X-rays, IRS~5 shows the most pronounced variability, whilst the other two class~I protostars IRS~1 and IRS~2 have more stable emission. X-ray data is also analyzed for the recently identified candidate class~0 source IRS~7E, which shows strong variability as well as for the Herbig~Ae star R~CrA for which we find extremely hot X-ray--emitting plasma. For IRS~1,2 and 5, the hydrogen column densities derived from the X-ray spectra are at about half the values derived with near-infrared techniques, a situation similar to what has been observed towards some other young stellar objects. 

   \keywords{Stars: pre-main sequence, Stars: individual: R CrA, 
   Radio continuum: stars, X-rays: stars}
   }

   \titlerunning{Radio and X-ray variability of YSOs in the \textsl{Coronet} cluster}

   \maketitle
%
%________________________________________________________________

\section{Introduction}

The energetic processes in and around pre-main sequence stars lead to
considerable variability over a wide spectral range. The study of
radio and X-ray emission allows to probe energetic processes in the coronae
of these stars (see \citealp{fem99}) in all their evolutionary stages: Low-mass stars evolve through a series 
of stages termed class~0 to class~III (\citealp{lad87}, and e.g.
\citealp{and00}). In the class~0 
stage, gravitational collapse has started, but most of the material is still
in the circumstellar envelope. Soon after the onset of accretion, powerful
bipolar outflows develop. The luminosity of class~I sources is still
mainly due to accretion, although a considerable fraction of the final matter
has already been accreted. The canonical age of such objects is assumed to be
about $10^5$ years. The class~I phase is followed by the class~II (=classical 
T Tauri star) and class~III (=weak-line T Tauri star) phases, during which the
circumstellar material plays a consecutively diminishing role in the
protostar's evolution.

\subsection{X-rays from young stellar objects}

X-ray emission from classical and weak-line T Tauri stars has been known
for some time and is thought to be produced in magnetically confined plasma
in the coronae \citep{fem99,fam03}. During the 1990s, X-ray emission
towards a few even younger class~I protostars was discovered (reviewed by
\citealp{neu97}). With the launches of \textsl{Chandra} and XMM-\textsl{Newton},
X-ray observatories with higher spatial resolution and better sensitivity
to harder photons (which are less affected by extinction)
became available, allowing better studies of star-forming regions, such as the
Orion Nebula Cluster \citep{gar00,fei02,fei03,get05}, $\rho$ Ophiuchi
\citep{ikt01,ogm05} or IC~348 \citep{prz01,prz02}. However, the
processes giving rise to magnetic activity in young stars in general and
class~0/I protostars in particular are still
poorly understood. For a review of the associated high-energy
processes, see \citet{fem99}, \citet{pre04r}, or, in the framework of stellar radio
and X-ray astronomy, \citet{gue02,gue04}. 

Many X-ray emitting class~I
protostars have X-ray luminosities which are more than ten times
higher than those of typical T Tauri stars while their X-ray spectra
are harder and show stronger absorption at lower energies due to circumstellar
extinction. Powerful flares have been observed frequently (e.g. \citealp{ikt01}).
Until now, however, only relatively few X-ray--detected class~I protostars
are known. A ROSAT survey of young stellar objects 
yielded only eleven class~I sources \citep{car98}. Searching for more X-ray
detections among class I protostars yields very different detection rates. 
While \citet{ikt01}, studying the $\rho$ Oph cloud, detect $\sim 70$ \% of
 the class~I
sources in X-rays, a result confirmed by \citet{ogm05}, \citet{pre04}
detects four out of 19 class~I sources in the Serpens dark cloud. 
\citet{ikt01} and \citet{pre04} suggest that class~I sources might be intrinsically more variable at X-ray wavelengths than protostars in more evolved evolutionary stages, i.e., class~II and class~III young stellar objects.

The first tentative X-ray detection of two candidate class~0 sources (in
OMC-3) was reported by \citet{tsu01}. However, the identification as
class~0 sources is ambiguous, \citet{tsu04} concluding that in one
case the X-ray emission is shock-induced by jets from a nearby class I source.
The detection of an X-ray flare from IRS~7E, a candidate class~0 source in the \textsl{Coronet} cluster, was first presented by \citet{ham05} and is discussed further in this paper.

When it comes to physical explanations for the hard X-ray emission from class~I protostars, often a magnetic interaction of the accretion disk and the stellar magnetosphere is put forward. In the model of \citet{hay96}, closed magnetic 
loops connecting the central star and the disk are twisted by 
the rotation of the disk (see also e.g. \citealp{shu97,bir00,iso03}).
The loops expand with increasing twist until reconnection takes
place, hot plasma giving rise to the X-ray and, conceivably,
nonthermal radio emission. This mechanism
predicts flare periodicity on the order of the rotation period of the
innermost parts of the disk and is compatible with observed flare
durations, i.e. several hours to days (e.g. \citealp{koy96}).
\citet{mon00}, basing their conclusions on the occurence of
quasi-periodic flares on YLW~15, propose an evolutionary scenario
explaining X-ray properties of protostars as due to their rotation
velocity. In this picture, young class~I sources in which the rotation
of star and disk is not yet synchronous, would constitute the
strongest X-ray emitting population of star
formation. This phase quickly
comes to an end (after a few $10^5$~years) and coronal activity
becomes the dominant source for X-ray emission. Thus, the following
part of the class~I stage is already comparable to the class~II phase
with synchronous rotation, however with much more extinction due to
circumstellar material than in the later evolutionary stages. Until
now, however, with no further periodicities found, observational
evidence for star-disk magnetic coupling in these objects is inconclusive
\citep{mon03}.
However, \citet{fav05} find evidence for such coupling from the analysis of
intense X-ray flares necessitating large magnetic structures.

\subsection{Radio emission from young stellar objects}

Thermal as well as nonthermal radio emission has been observed towards pre-main
sequence stars. Distinctive features of nonthermal radio emission include rapid
variability, circular polarization, negative spectral indices and exceedingly
high brightness temperatures (at high spatial resolution). Identified as
gyrosynchrotron emission from mildly relativistic electrons gyrating in
magnetic fields in this context, nonthermal radio emission is a 
direct tracer of circumstellar magnetic activity. Thermal radio emission
is thought to be produced by ionized material e.g. at the base of outflows.
The radio properties of young stellar objects
are reviewed by e.g. \citet{and96}; for a discussion of gyrosynchrotron
radio emission from a young B star, see \citet{and88}.

A number of X-ray emitting classical and weak-line T Tauri stars are
sources of nonthermal radio emission produced 
in large-scale magnetic structures in the coronae of these objects 
also giving rise to 
thermal X-ray emission from coronal plasmae. In particular,
circularly polarized radio emission indicative of
gyrosynchrotron emission has been observed towards several
YSOs (e.g. \citealp{and88,and92,whi92,ski93,phi93,rod99}). 

Class~I protostars have mostly been observed to be thermal radio
sources. \citet{sha04} model the centimetric free-free emission of 
class~I protostars based on the X-wind model \citep{shu97} with X-ray 
emission as the main ionization source, and indeed some of the class~I 
sources showing X-ray emission are known radio sources as well (e.g.
\citealp{gro97,fcw98,sgb99}). However, already at the base of such outflows,
the material becomes optically thick so that emission from the immediate 
surroundings of the star -- e.g. nonthermal radiation caused by magnetic
fields -- can easily be hidden. \citet{gib99} proposes a scenario where only 
the earliest and latest stages of star formation emit at radio
wavelengths: With fading accretion, the (thermal) jet emission decreases while
at the same time presumably a stellar magnetic field builds up,
giving rise to nonthermal emission. 

As at X-ray wavelengths, not all class I protostars are detected at radio wavelengths either: \citet{luc00} detect four out of seven
class~I sources with outflow activity in Taurus, while \citet{leh03}
detect one out of four associated to Cederblad 110, interestingly the
one with the largest far-infrared flux. These low numbers remain
inconclusive concerning the fraction of radio-detected class~I
sources but might still point towards an evolutionary dependence
within the class~I stage.

Among the three class~I sources discussed here in detail, one (IRS~5) is 
observed to emit nonthermal radio emission (as reported earlier by 
\citealp{fcw98}). This source is in fact the only class~I protostar
known to emit such radiation. Already its circular polarization is a clear sign, additionally its short-term variability is difficult or impossible
to reconcile with thermal emission.

At present, little is known about the temporal correlation of the X-ray and 
radio emission, especially when nonthermal radio emission is concerned. Here,
we study these phenomena separately, based on non-simultaneous observations.
The only possibility to learn more about these sources consists of 
simultaneous multi-wavelength studies looking
for correlations in the variability at different wavelengths, especially in 
the radio and X-ray regimes. Until now, only few systematic attempts of such 
observations of young stellar objects have been made: \citet{fei94} observed
a magnetically active T Tauri star, \citet{gue00} targeted pre-main sequence 
stars in the Taurus-Auriga region, specifically V773~Tau, and \citet{gag04}
observed pre-main sequence stars in the $\rho$ Oph cloud complex. However, only
in the latter case two class~I protostars were targeted but not detected at 
both radio and X-ray wavelengths. Simultaneous observations like these can 
be very valuable, as underlined by the serendipitous discovery of an 
enormous flare at radio and X-ray wavelengths, apparently from a highly 
obscured weak-line T Tauri star in Orion \citep{bow03}.

\subsection{Protostars in the \textsl{Coronet} cluster}

\begin{figure*}
 \centering
 \includegraphics[width=6.2cm, angle=-90]{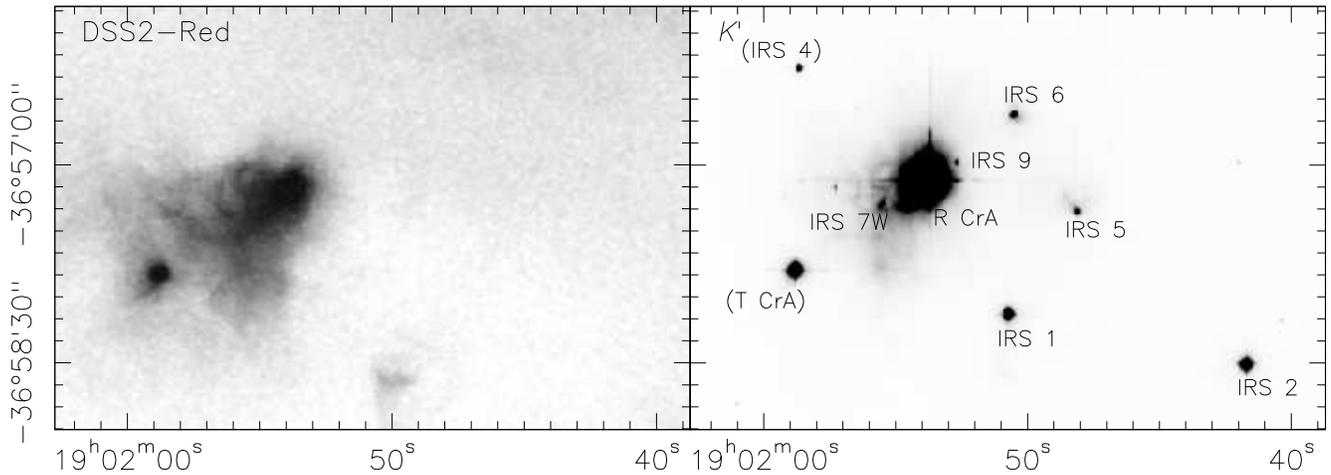}
 \caption{Optical (DSS2-Red, left) and near-infrared ($K'$, right) views of the \textsl{Coronet}
 cluster. T~CrA, seen at left, is not covered by the radio data presented here.
 It is marginally detected in X-rays \citep{ski04}. The $K'$ image is from
 \citet{hod94}, coordinates are J2000.}
\label{irdssview}
\end{figure*}

\begin{figure*}
%\sidecaption
\centering
 \includegraphics[width=11cm]{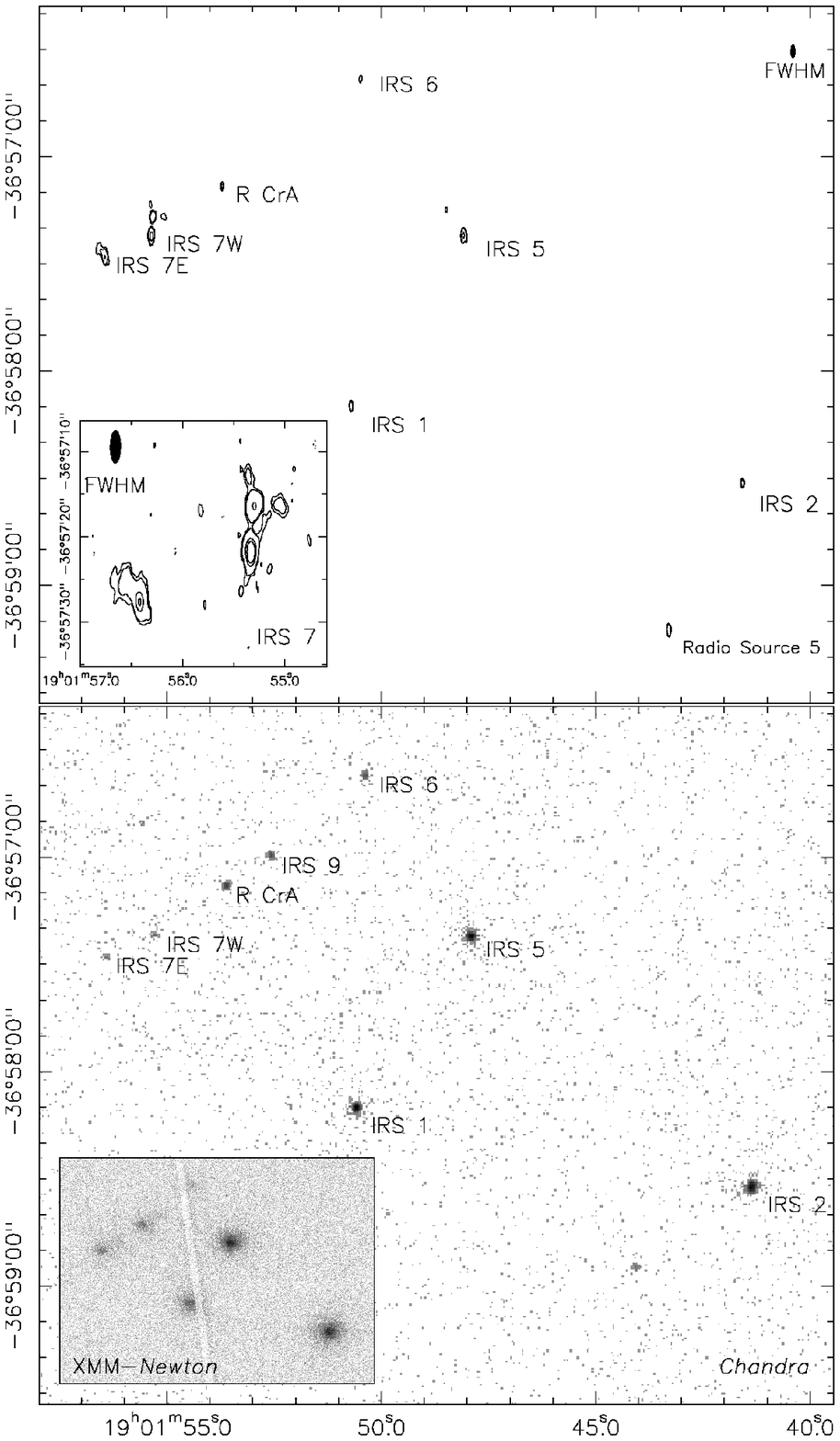}
  \caption{ Upper panel: Image made from the combined $uv$ data of all five VLA observation epochs. Contours shown are 5, 10 and 100 $\sigma$ in the main plot and --3 (dashed),3,5,50 and 100 $\sigma$ in the inset showing IRS~7. Coordinates
are J2000.0. Lower panel: The same view, showing the merged \textsl{Chandra} ACIS-I data. The inset shows approximately the same view in the merged XMM-\textsl{Newton} data (MOS+EPN, X1-X3).}
\label{chandraview}
\label{vlainset}
\end{figure*}

The Corona Australis (CrA) molecular cloud complex is a nearby region
with ongoing star formation, at a distance of only 150~pc, which
is similar to that of the Taurus and $\rho$ Oph dark cloud complexes
\citep{fim84}\footnote{Literature values for the distance range from 130~pc
\citep{mar81} to 170~pc \citep{knh98}.}. \citet{tas84} found a compact
cluster of infrared sources in the densest part of the cloud complex
close to the Herbig Ae star R~CrA, calling it the
\textsl{Coronet} cluster. Fig.~\ref{irdssview} shows the DSS2-Red and $K'$-band
views of its central region, clearly showing a number of deeply
embedded sources.

The \textsl{Coronet} cluster is an association of young stellar objects
mostly identified at radio as well as X-ray wavelengths. It is comprised of
at least five deeply embedded class~I protostars
\citep{wil86,wil92,wil97}, four of which are covered by the
observations presented here\footnote{The fifth one, IRS~3, is located
outside the VLA primary beam area and remains undetected at X-ray
energies (cf. \citealp{nis05}).}. \citet{koy96} reported the detection of hard X-ray emission
towards these five protostars. Three of these sources
could be verified at higher angular resolution by ROSAT
\citep{nep97}, more ROSAT-detected young stellar objects were
reported by \citet{neu00}. First results from \textsl{Chandra}-ACIS
observations of this region were presented by \citet{gar03} who
detected more than 70 sources in a 20~ksec exposure.

The \textsl{Coronet} cluster has been analyzed at radio wavelengths
already several times. \citet{bro87} carried out VLA observations at
$\lambda=6$~cm in 1985, identifying five sources associated with young stellar
objects. \citet{sut96} analyze the radio variability of
pre-main-sequence stars using archival VLA (1985-1987) and Australia
Telescope (1992) data, while \citet{fcw98} report on VLA observations 
at 3.5~cm, taken in 1997. They detect circular polarization toward the 
class~I protostar IRS~5.

Studying dust emission from the region at $\lambda = 1.2$ mm,
\citet{chi03} report several peaks, with the brightest one (called MMS~13
and already detected by \citealp{hen94})
in the vicinity of IRS~7 in the \textsl{Coronet} cluster. Millimeter emission
is associated with IRS~1, 2, and 5 as well. They conclude that the MMS~13 peak
probably is a class~0 source not related to any of the sources observed at
other wavelengths. Additionally based on millimeter molecular line data,
\citet{gro04} reach a similar conclusion. 
\citet{nut05} present higher-resolution submillimeter continuum data
resolving MMS~13 into two protostellar and one prestellar source where
gravitational collapse has not yet started. This prestellar source is at
the position of the class~0 source surmised earlier and is estimated to 
have a mass of 6 to 11~M$_\odot$, while the envelopes of the protostars 
considered in this work all have sub-solar mass dust masses. We note already
here that the two protostellar sources are associated with unresolved radio
and X-ray sources.

Using near-infrared spectroscopy, \citet{nis05}
constrain the accretion rates and evolutionary stages of some
\textsl{Coronet} protostars. They conclude that IRS~1 and IRS~2 appear
to be mostly accretion-powered (accounting for 80\% to 65\% of L$_{\rm
bol}$, respectively), while IRS~5 -- a binary source -- appears to be a
deeply embedded, more evolved object with accretion
accounting for only 20\% of the bolometric luminosity of one of the components,
IRS~5a. IRS~3 and IRS~6 show no signs of accretion. Comparing their
derived stellar luminosities and effective temperatures to different
models of pre-main sequence evolution, IRS~1 emerges as the youngest
object with an age of about $10^5$ years, while IRS~2, 5a and 6a
apparently have about the same age, closer to $10^6$ years in spite of their
different accretion properties.

\section{Observations and Data Analysis}
\subsection{Radio observations}
Here we present the results of a radio monitoring campaign
at  $\nu=8.44$ GHz of the
\textsl{Coronet} cluster, carried out with the NRAO Very Large Array
(VLA) between 1998 September 19 and October 13.
The sources, which are all within the 5.4 arcmin FWHM primary beam,
were observed five
times, separated by a few days each.
Originally, these observations
were scheduled to be simultaneous with X-ray monitoring, but ROSAT
unexpectedly went out of operation
before the campaign was to begin. The
VLA was observing in the B configuration (program ID AM596) at $\nu=8.44$ GHz
with two intermediate frequency (IF) pairs, offset by each IF band 's width,
50 MHz. One IF of each pair detected right circular polarization (RCP),
the other one left circular polarization (LCP).
 The phase center was $(\alpha, \delta)_{ {\rm
J}2000}= 19^{\rm h}$01$^{\rm m}$48$^{\rm s}, -36^\circ$57'59".
These data are complemented by four additional epochs in 1998, also taken
with the VLA in B/BnA configuration (program ID AK469) at $\nu=8.44$ GHz, increasing the total to nine epochs, now starting on June 27th, 1998.
The source coverage is slightly different due to a different phase
center, $(\alpha, \delta)_{ {\rm J}2000}= 19^{\rm h}$01$^{\rm m}$55.6$^{\rm s},
-36^\circ$57'09.6".
Details on the dates and durations of the different observation epochs are given in
Table~\ref{obslist}. These data can be compared with measurements taken by
\citet{fcw98} on 1997 January 19 and 20 (see Table~\ref{sourceralist}).

\begin{table*}
%\begin{center}
\caption[]{VLA \textsl{Coronet} Observation Dates}
\begin{tabular}{llllll}
\hline
Epoch & Day & Date & IAT Interval & Duration & rms ($\mu$Jy) \\
\hline
\hline
R1 &  1  & 1998 Jun 27 & 07:21:30 -- 08:50:20     & 01:28:50 & 30\\
R2 & 23  & 1998 Jul 19 & 05:27:10 -- 06:54:00     & 01:26:50 & 33\\
R3 & 73  & 1998 Sep 07 & 03:08:30 -- 04:07:21     & 00:58:51 & 36\\
R4 & 85  & 1998 Sep 19 & 01:23:00 -- 03:17:40     & 01:54:40 & 24\\
R5 & 93  & 1998 Sep 27 & 01:51:40 -- 03:46:10     & 01:54:30 & 25\\
R6 & 98  & 1998 Oct 02 & 00:03:00 -- 01:56:40     & 01:53:40 & 27\\
R7 &102  & 1998 Oct 06 & 00:15:11 -- 01:13:10     & 00:57:59 & 36\\
R8 &106  & 1998 Oct 10 & 23:52:50 -- 01:51:10(+1) & 01:58:20 & 23\\
R9 &109  & 1998 Oct 13 & 23:14:40 -- 01:09:30(+1) & 01:54:50 & 25\\
\hline
\label{obslist}
\end{tabular}

All epochs except for R1 (BnA) were observed with the B array.\\
Datasets R4-R6 and R8-R9 are from project AM596\\
Datasets R1-R3 and R7 are from project AK469\\
%\end{center}
\end{table*}

The VLA data were analyzed using NRAO's Astronomical
Image Processing System (AIPS).
An absolute flux density scale was established by observations
of 3C286, the nearby phase calibrator
1924$-$292 was observed every ten minutes. 
At a declination of nearly $-37^\circ$, the \textsl{Coronet}
cluster is observable with the VLA only at very low elevations
($<18^\circ$), increasing problems with phase instabilities caused by
the atmosphere. Here, however, by using self-calibration, these
effects could be minimized. All observation epochs were
phase-only self-calibrated, as the low flux densities did not
warrant self calibration of, both, phases plus amplitudes. Following this, 
Stokes-$I$ and Stokes-$V$ were imaged separately, with the
input $uv$-data split into different time intervals.
The polarization of the secondary calibrator 1924$-$292 is $\lesssim 1$\%.
Source positions and fluxes were determined, using the
AIPS task SAD, by fitting Gaussians to each detection. The light curves
thus derived are shown in Fig.~\ref{vlacurves}. The root mean square (rms) 
noise level in the Stokes $I$ and $V$ maps for the different epochs is
shown in Table\ref{obslist}. For the detection of weak sources, the
five AM596 observations were combined, yielding a map with an rms of
$<12\mu$Jy. The synthesized beam size for this combined data is 
$1.93 \times 0.63$'' at position angle $-0.51^\circ$.

\subsection{Archival X-ray data}
In addition to the radio data, we studied archival X-ray data of the 
\textsl{Coronet} sources. On 2000 October 7, this region was observed with
\textsl{Chandra}, using the Advanced CCD Imaging
Spectrometer in its imaging mode ACIS-I in order to search for X-rays
from young stars in a 20~ksec exposure \citep{gar03}. A second
ACIS-I observation of this region, with a 38~ksec
exposure, was carried out on 2003 June 26. Additionally, there have
been three XMM-\textsl{Newton} observations of the \textsl{Coronet}
region: On 2001 April 9 there was a 26~ksec exposure and on
2003 March 28/29 two consecutive observations of 36~ksec and 32~ksec 
were carried out. While the first XMM-\textsl{Newton} observation was
taken through the ``thin'' filter, the ``medium'' filter was used for the 
latter two observations. Details on the X-ray data analyzed here are
listed in Table~\ref{xraylist}.
Our analysis of the archival X-ray data focuses on the three
X-ray brightest sources among the sample discussed here, IRS~1,2, and
5 because they are bright enough for a spectral analysis. These three
sources are class~I protostars.

\begin{table*}
%\begin{center}
\caption[]{Archival X-ray data used in this study}
\begin{tabular}{llrlllll}
\hline
& Satellite & Obs. & ID & Date & Time & Duration & Pointing Center\\
&          &      &    &      & (UT) & (sec)   & R.A./Dec. (J2000)\\
\hline
\hline
1&\textsl{Chandra}    &         19 & C1 & 2000 Oct 07 & 17:01:59 --
23:11:01     & 19958 & 19 01 50.6 -36 57 30.0\\
2&XMM-\textsl{Newton} & 0111120101 & X1 & 2001 Apr 09 & 09:32:16 --
15:56:46     & 26445 & 19 01 49.4 -36 55 41.4\\
3&XMM-\textsl{Newton} & 0146390101 & X2 & 2003 Mar 28 & 08:50:45 --
18:32:49     & 36715 & 19 01 37.2 -36 51 02.3\\
4&XMM-\textsl{Newton} & 0146390201 & X3 & 2003 Mar 29 & 19:35:00 --
04:00:38(+1) & 32211 & 19 01 37.2 -36 51 02.5\\
5&\textsl{Chandra}    &       3499 & C2 & 2003 Jun 26 & 12:58:10 --
00:16:08(+1) & 38126 & 19 01 50.6 -36 57 30.0\\
%Roll Angles: C1: 276.01, C2: 36.97, X1: 86.62, X2/X3: 97.73 (C:Roll
%Angle, X: Actual Position Angle)
\hline
\label{xraylist}
\end{tabular}
%\end{center}
\end{table*}

\textsl{Chandra} X-ray data were analyzed using the Chandra Interactive
Analysis of Observations (CIAO) 3.1 software package together with 
CALDB 2.28. 
Apertures used were three or five (IRS~1,2, and 5)
arcseconds in diameter. Spectra were prepared with the task 
\textsl{psextract}, including the creation of response matrices and 
ancillary response files. Finally, spectra bin sizes were determined by
a minimum event number per bin of fifteen.

XMM-\textsl{Newton} data were analyzed using the Science Analysis System
(SAS) 6.0.0. Due to the superior signal-to-noise ratio among XMM imaging
instruments, we only make use of the EPIC-PN data here.
We generally used apertures with a diameter of 32'', but in cases of close neighbouring sources, we defined smaller (14'' diameter) non-overlapping extraction regions.
These apertures encircle approximately 66\% and 40\% of
the point source flux\footnote{as given in the XMM-\textsl{Newton} User's
Handbook V2.2}, so that corresponding correction factors had to be taken
into account in the luminosity determination. Spectroscopy involved the 
creation of response matrices and ancillary
response files using the tasks \textsl{rmfgen} and \textsl{arfgen}, 
respectively. Finally, the spectra were binned with the same minimum 
event number criterion as was used for the \textsl{Chandra} data.

Models were fitted to all background-subtracted spectra with
\textsl{Sherpa} from the CIAO 3.1 software package, using Monte Carlo
multi-parameter fits. The spectral model consists of the Astrophysical Plasma
Emission Code (APEC) with an absorption factor. This code calculates the
emission spectrum from collisionally ionized diffuse gas. The main fit
parameters are the plasma temperature, the absorbing hydrogen column 
density and the element abundances compared to solar values.

After a determination of
elemental abundances from the Fe K line at 6.7~keV from one spectrum
per source, an absorbed APEC emission model was fitted to the
spectra. It turned out that the spectra can be well described with a
single-temperature model; there was no need to assume two temperature
components. However, due to the high absorbing column densities
toward the sources, softer emission components might easily be
extincted. The uncertainties of the best-fit parameter values 
were estimated with the \textsl{Sherpa} task 'uncertainty'. Based on
these fits, the count rates in the lightcurves were converted into unabsorbed 
flux using the PIMMS software \citep{muk93}
and finally, into luminosity units (energy range 0.5 -- 10 keV), assuming
a distance of $d=150$~pc.

\section{Results}

After a general discussion of the 1998 VLA monitoring and the
analysis of archival X-ray data, individual source properties will be
summarized.

\begin{table}
%\begin{center}
\caption{Positions and identification of \textsl{Coronet} radio and X-ray sources}
\begin{tabular}{lll}
\hline
No. & Position (J2000.0) & Source ID$^{\rm a}$ \\
\hline
\hline
1  & 19 01 41.6 -36 58 31 & IRS~2 (I)          \\
2  & 19 01 43.3 -36 59 12 & Source 5$^{\rm b}$ \\
3  & 19 01 44.2 -36 58 54 &                    \\ 
4  & 19 01 48.1 -36 57 22 & IRS~5 (I)          \\ 
5  & 19 01 48.5 -36 57 15 &                    \\
6  & 19 01 50.5 -36 56 38 & IRS~6 (T)          \\
7  & 19 01 50.7 -36 58 09 & IRS~1 (I)          \\
8  & 19 01 52.6 -36 57 01 & IRS~9 (I)          \\
9  & 19 01 53.7 -36 57 08 & R CrA (H)          \\
10 & 19 01 55.0 -36 57 16 &                    \\
11 & 19 01 55.3 -36 57 16 & Source 9$^{\rm b}$ \\
12 & 19 01 55.3 -36 57 22 & IRS~7W             \\
13 & 19 01 55.4 -36 57 13 &                    \\
14 & 19 01 56.4 -36 57 28 & IRS~7E (0?)         \\
15 & 19 01 56.5 -36 57 26 &                    \\
\hline
\label{sourceidlist}
\end{tabular}

$^{\rm a}$ types: (0)=class~0, (I)=class~I, (T)=T Tau, (H)=HAe\\
$^{\rm b}$ as defined by \citet{bro87} \\
\end{table}

\begin{table*}
%Quelle: AK469.sxc und sad30min.sxc
%VLA flux values are phase-only self-cal.
%\begin{center}
\caption{\textsl{Coronet} sources, radio data}
\begin{tabular}{lrrrrrrrrrrr}
\hline
No. &  R1    & R2    & R3    & R4   & R5   & R6   & R7    & R8   & R9    & combined$^{\rm a}$ & FCW98$^{\rm b}$ \\
    & (mJy) & (mJy) & (mJy) & (mJy)& (mJy)& (mJy)& (mJy) & (mJy)& (mJy) & (mJy)              & (mJy) \\ 
\hline
\hline
1  &  0.30  &$<0.27^{\rm c}$&$<0.30^{\rm c}$& 0.29 & 0.35 & 0.32 &$<0.30^{\rm c}$& 0.36 & 0.30  & --		     & 0.67 \\
2  &  0.40  & 0.68  & 1.15  & 1.04 & 0.76 & 0.93 & 0.52  & 1.21 & 0.58  & --		     & 1.08 \\
3  &  n.d.  & n.d.  & n.d.  & n.d. & n.d. & n.d. & n.d.  & n.d. & n.d.  & --		     & n.d. \\ 
4  &  1.96  & 1.10  & 0.75  & 1.32 & 1.11 & 2.97 & 2.31  & 3.28 & 1.67  & --		     & 1.36 \\ 
   & $<0.09^{\rm c}$& 0.22  & 0.15  & 0.25 & 0.26 & 0.29 & 0.34  & 0.65 & 0.36  & --         & 0.23 \\
5  &  n.d.  & n.d.  & n.d.  & n.d. & n.d. & n.d. & n.d.  & n.d. & n.d.  & 0.10  	     & n.d. \\
6  &  n.d.  & n.d.  & n.d.  & n.d. & n.d. & n.d. & n.d.  & n.d. & n.d.  & 0.15  	     & n.d. \\
7  &  0.42  & 0.36  & 0.59  & 0.44 & 0.47 & 0.45 & 0.55  & 0.49 & 0.46  & --		     & 0.45 \\
8  &  n.d.  & n.d.  & n.d.  & n.d. & n.d. & n.d. & n.d.  & n.d. & n.d.  & $<0.035^{\rm c}$   & n.d. \\
9  &  0.16  & 0.21  & 0.17  & 0.22 & 0.17 & 0.23 & 0.24  & 0.20 & 0.26  & --		     & 0.23 \\
10 &  n.d.  & n.d.  & n.d.  & n.d. & n.d. & n.d. & n.d.  & n.d. & n.d.  & 0.12  	     & n.d. \\
11 &  0.93  & 0.78  & 0.89  & 0.85 & 0.78 & 0.72 & 1.04  & 0.94 & 0.84  & --		     & 1.58 \\
12 &  4.26  & 4.15  & 5.01  & 4.27 & 4.31 & 4.11 & 5.29  & 4.52 & 4.31  & --		     & 3.74 \\
13 &  n.d.  & n.d.  & n.d.  & n.d. & n.d. & n.d. & n.d.  & n.d. & n.d.  & 0.11  	     & n.d. \\
14 &  1.38  & 1.53  & 2.10  & 1.66 & 1.83 & 1.54 & 2.04  & 1.75 & 1.62  & --		     & 2.03 \\
15 &  n.d.  & n.d.  & n.d.  & n.d. & n.d. & n.d. & n.d.  & n.d. & n.d.  & 0.13  	     & n.d. \\
\hline
\label{sourceralist}
\end{tabular}
%\end{center}

$^{\rm a}$ sources detected in combined AM596 data (10h) \\
$^{\rm b}$ 3.5~cm flux on 19/20 Jan 1997, as published by \citet{fcw98}\\
$^{\rm c}$ local 3$\sigma$ limit, in the case of IRS~2: accounting for primary beam attenuation \\
\end{table*}

\begin{table*}
%\begin{center}
\caption{\textsl{Coronet} sources, X-ray-data}
\begin{tabular}{ll|rr|rrr}
\hline
No. & Source ID & C1$^{\rm a}$           & C2$^{\rm a}$    & X1$^{\rm a}$          & X2$^{\rm a}$            & X3$^{\rm a}$ \\
    &           & (cnts/ksec)            & (cnts/ksec)     & (cnts/ksec)           & (cnts/ksec)             & (cnts/ksec)  \\
\hline
\hline
%.                C1                       C2                X1                      X2                        X3  
1  & IRS~2      & $48.7 \pm 1.6$         & $26.6 \pm 0.9$  & $42.1 \pm 1.5$        & $49.4 \pm 1.4$          & $54.4 \pm 1.5$ \\
2  & Source 5   & n.d.                   & n.d.            & n.d.                  & n.d.                    & n.d.	         \\
3  &            & $0.5  \pm 0.2$         & $0.8  \pm 0.2$  & n.d. 	           & n.d. 	             & n.d.	       \\ 
4  & IRS~5      & $13.2 \pm 0.9$         & $31.0 \pm 0.9$  & $46.5 \pm 1.6$        & $40.3 \pm 1.3$          & $75.6 \pm 1.7$\\ 
5  &            & n.d. 		         & n.d.		   & n.d.	           & n.d.	             & n.d.	       \\ 
6  & IRS~6      & $ 2.4 \pm 0.4$         & $ 1.5 \pm 0.2$  & $ 3.4 \pm 0.8$        & $ 0.9^{\rm c} \pm 0.5$  & $ 0.5^{\rm c} \pm 0.5$\\ 
7  & IRS~1      & $22.9 \pm 1.1$         & $30.9 \pm 0.9$  & $31.7 \pm 1.4$        & $10.9^{\rm c} \pm 0.7$  & $18.3^{\rm c} \pm 0.9$\\
8  & IRS~9      & $ 2.9^{\rm b} \pm 0.4$ & $2.1 \pm 0.3$   & $1.3^{\rm d} \pm 0.4$ & $2.8^{\rm d} \pm 0.4$   &  $1.2^{\rm d} \pm 0.4$\\
9  & R CrA      & $ 3.2 \pm 0.5$         & $ 3.1 \pm 0.3$  & $5.2^{\rm d} \pm 0.6$ & $ 9.3^{\rm d} \pm 0.6$  & $ 6.8^{\rm d} \pm 0.6$\\
10 &            & n.d.		         & n.d.		   & n.d.	           & n.d.	             & n.d.	       \\
11 & Source 9   & n.d. 		         & n.d. 	   & n.d.	           & n.d.	             & n.d.	       \\
12 & IRS~7W     & $ 0.6 \pm 0.2$         & $ 0.6 \pm 0.2$  & $2.1^{\rm d} \pm 0.4$ & $ 2.6^{\rm d} \pm 0.4$  & $ 4.7^{\rm d} \pm 0.5$\\
13 &            & n.d.		         & n.d.		   & n.d.	           & n.d.	             & n.d.	       \\
14 & IRS~7E     & $ 1.0 \pm 0.3$         & $ 0.3 \pm 0.1$  & $1.6^{\rm d} \pm 0.4$ & $ 3.7^{\rm d} \pm 0.4$  & $14.7^{\rm d} \pm 0.8$\\
15 &            & n.d.		         & n.d.		   & n.d.	           & n.d.	             & n.d.	       \\
%IRS9, C1, flare $^{\rm f}$, var 6
\hline
\label{sourcexrlist}
\end{tabular}
%\end{center}

$^{\rm a}$ background-subtracted count rates (0.5 -- 10 keV) from X-ray datasets as listed in Table~\ref{xraylist}, \textsl{Chandra} and XMM-\textsl{Newton} count rates are not directly comparable. Additionally, different filters were used in the XMM-\textsl{Newton} observations: 'thin' for X1 and 'medium' for X2/X3. \\
$^{\rm b}$ flare \\
$^{\rm c}$ source partly covered by chip gap, loss estimated to be $\approx 20$\% \\
$^{\rm d}$ contamination from neighbouring source \\
\end{table*}

\begin{table*}
%1sigma from irs*_1s.dat
%\begin{center}
\caption{X-ray spectral fitting results for IRS 2, 1, 5, and R CrA: Absorbing column density and plasma temperature, with $1\sigma$ error margins}
\begin{tabular}{lllllllll}
\hline
   & IRS 2                    &		             & IRS 5	                &		       & IRS 1		          &	                 & R CrA	          &	\\
   & $N_H$                    & $T$  	             & $N_H$	                & $T$  	               & $N_H$	                  & $T$  	         & $N_H$	          & $T$  	\\
   & [10$^{22}$~cm$^{-2}$]    & [MK]                 & [10$^{22}$~cm$^{-2}$]    & [MK]                 & [10$^{22}$~cm$^{-2}$]    & [MK]                 & [10$^{22}$~cm$^{-2}$]  & [MK]\\
\hline
\hline
C1 & $2.43^{+0.11}_{-0.10}$   & $45.9^{+3.3}_{-3.3}$ & $3.02^{+0.25}_{-0.22}$   & $26.7^{+2.5}_{-2.3}$ & $3.75^{+0.31}_{-0.27}$   & $63.1^{+9.2}_{-6.6}$ &  --  			& --\\
X1 & $1.99^{+0.09}_{-0.08}$   & $27.6^{+1.7}_{-1.7}$ & $3.50^{+0.15}_{-0.13}$   & $30.2^{+1.2}_{-1.2}$ & $4.60^{+0.23}_{-0.21}$   & $28.7^{+1.4}_{-1.4}$ &  $1.2^{+0.4}_{-0.3}$   & $106^{+209}_{-43}$ \\
X2 & $1.76^{+0.06}_{-0.06}$   & $28.8^{+1.4}_{-1.4}$ & $4.80^{+0.17}_{-0.15}$   & $26.1^{+0.9}_{-0.9}$ & $3.21^{+0.31}_{-0.26}$   & $48.9^{+7.7}_{-6.0}$ &  $1.4^{+0.2}_{-0.2}$   & $100^{+57}_{-26}$ \\
X3 & $1.60^{+0.15}_{-0.16}$   & $36.5^{+5.7}_{-4.2}$ & $4.23^{+0.13}_{-0.12}$   & $36.8^{+1.1}_{-1.1}$ & $2.75^{+0.21}_{-0.18}$   & $44.6^{+4.4}_{-4.3}$ &  $1.3^{\rm a}      $   & $110^{+145}_{-45}$ \\
C2 & $1.53^{+0.07}_{-0.07}$   & $37.5^{+2.7}_{-2.6}$ & $4.07^{+0.16}_{-0.14}$   & $33.3^{+1.4}_{-1.4}$ & $2.36^{+0.09}_{-0.09}$   & $32.6^{+1.7}_{-1.6}$ &  --  			& --\\
\hline
\label{fitrestab}
\end{tabular}

$^{\rm a}$ frozen parameter, otherwise $N_H = 2.6^{+0.5}_{-0.4} \times 10^{22}$~cm$^{-2}$ , $T = 50^{+14}_{-9}$~MK; however this
change appears unlikely since X3 is immediately following X2
%\end{center}
\end{table*}

\subsection{Radio Monitoring}

In Stokes-$I$, eight sources can be detected throughout all five AM596
observation epochs, taking a 5$\sigma$ detection limit
as the criterion for source existence. These sources are also detected
throughout the AK469 data with the exception of IRS~2, not detected
due to primary beam attenuation far from the observation's phase center.
An image created from the combined AM596 $uv$ data is shown in Fig.~\ref{vlainset}.
Five weak additional sources only emerge from this
integration of the five AM596 observation epochs, including the T Tauri star IRS~6.
In Stokes-$V$ only IRS~5 was detected, showing epoch-averaged circular polarization
degrees between 0\% and 23\%. IRS~5 at the same time is the most
spectacularly variable source in this dataset: Starting at a relatively high emission level,
but interestingly without circular polarization, the source dims, then rises sharply
towards two peaks at the end of the period covered, accompanied by changes in its polarization.
Analyzing the data in shorter intervals, the Stokes-$V$ emission of
IRS~5 remains variable at timescales of 0.5 h while this is not the
case for any of the sources in Stokes-$I$. 
The remaining sources show diverse degrees of variability, but to a lesser extent
(Fig.~\ref{vlacurves}). All sources detected in single epochs have been 
detected at radio wavelengths before. The source fluxes derived from the VLA
observations are summarized in Table~\ref{sourceralist},
together with the 3.5~cm fluxes from \citet{fcw98} for comparison.

\subsection{Archival X-ray data}

\begin{figure*}
\begin{minipage}{3.5cm}
      \centerline{\hbox{
      \includegraphics*[width=3.9cm]{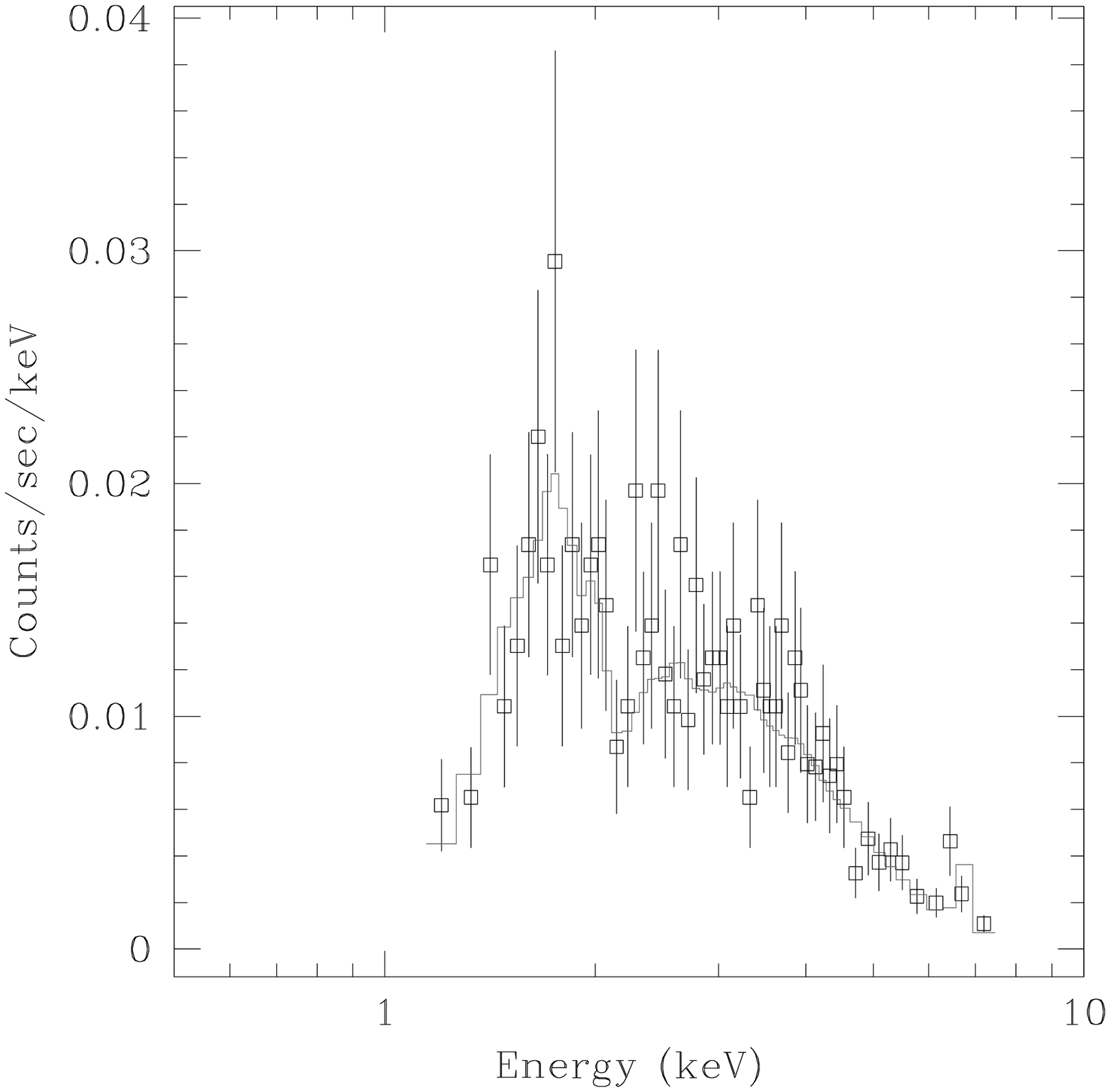}
      }}
      \centerline{IRS 2 : C1}
\end{minipage}
\begin{minipage}{3.5cm}
      \centerline{\hbox{
      \includegraphics*[width=3.9cm]{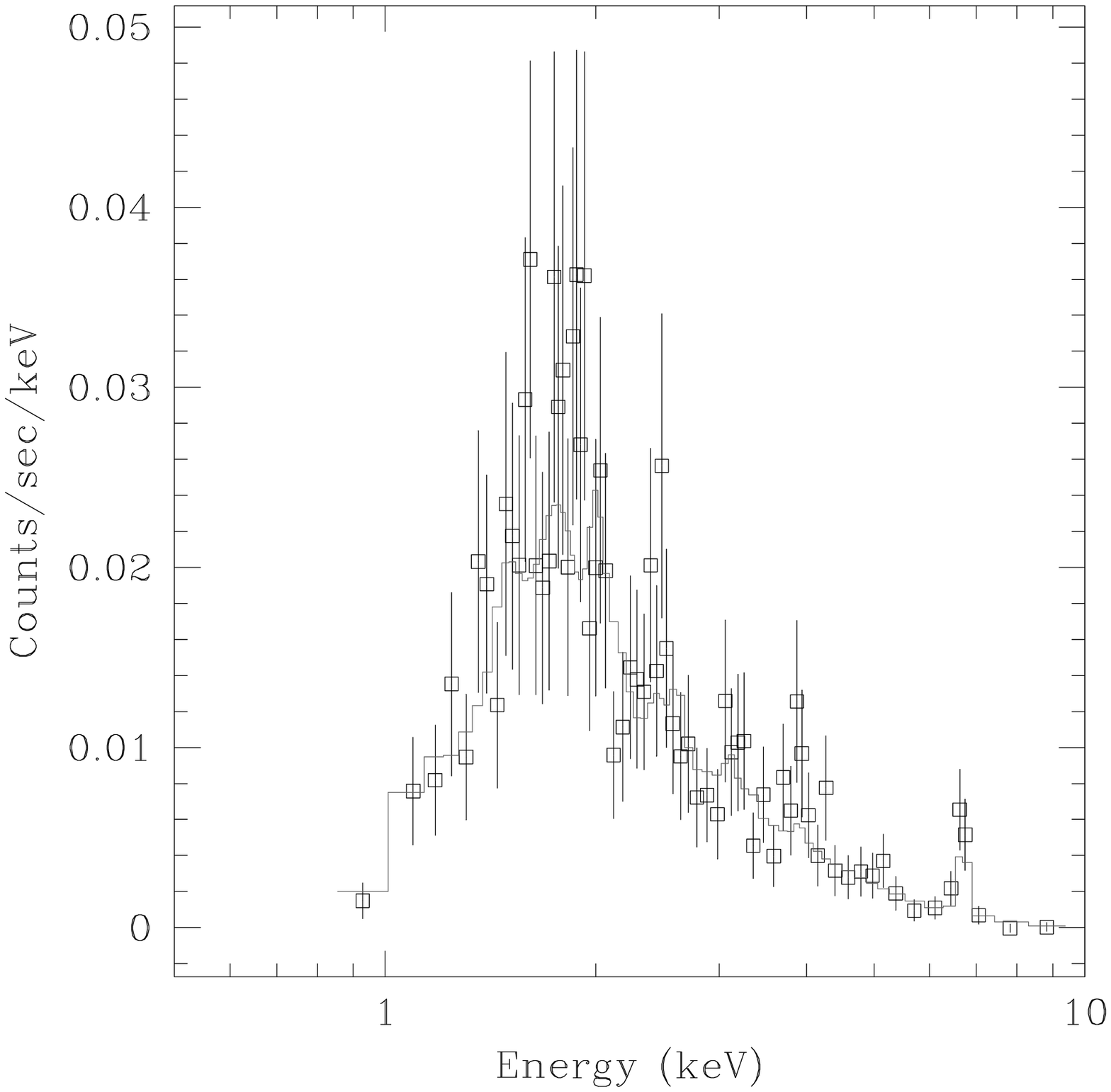}
      }}
      \centerline{X1}
\end{minipage}
\begin{minipage}{3.5cm}
      \centerline{\hbox{
      \includegraphics*[width=3.9cm]{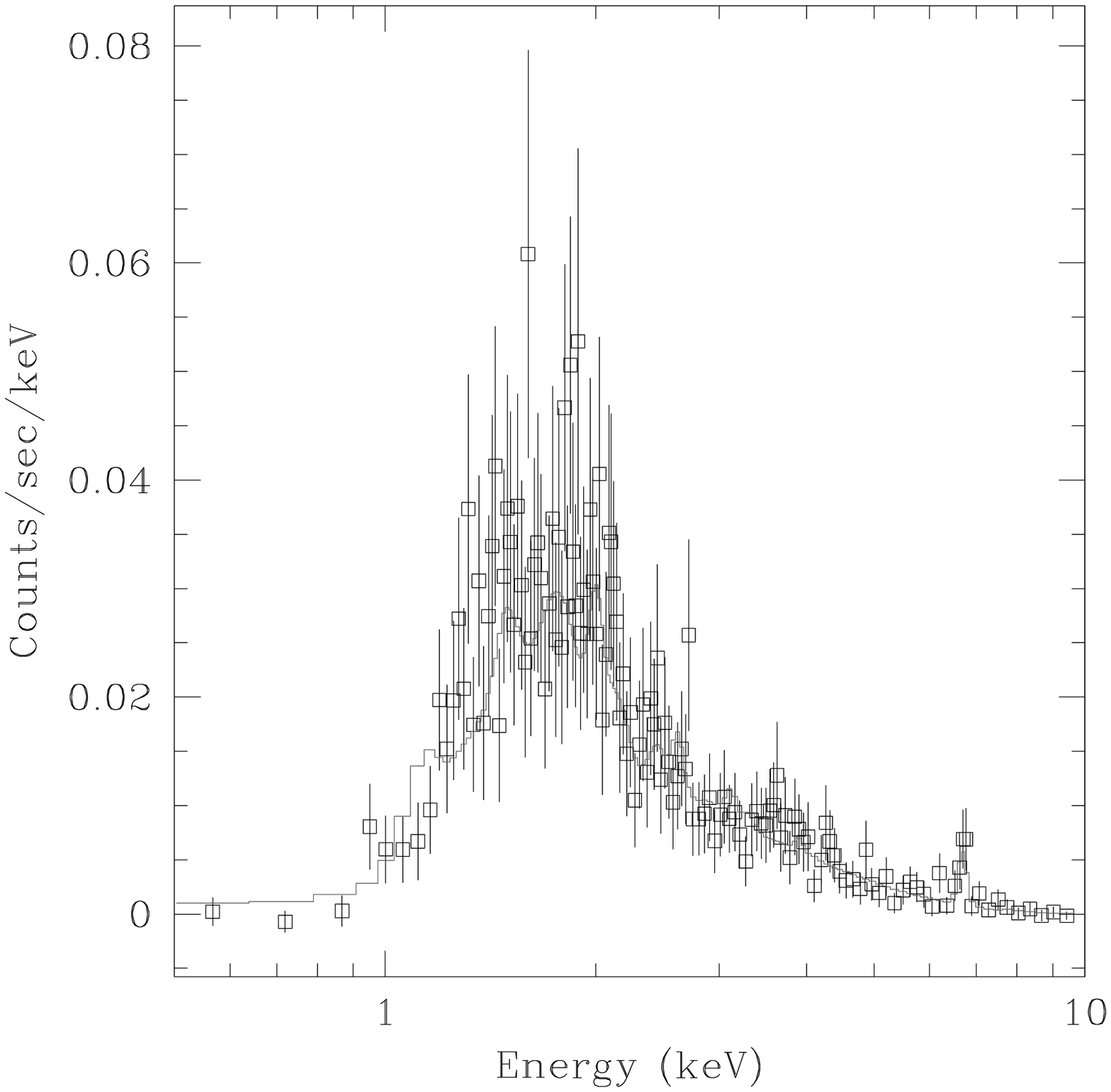}
      }}
      \centerline{X2}
\end{minipage}
\begin{minipage}{3.5cm}
      \centerline{\hbox{
      \includegraphics*[width=3.9cm]{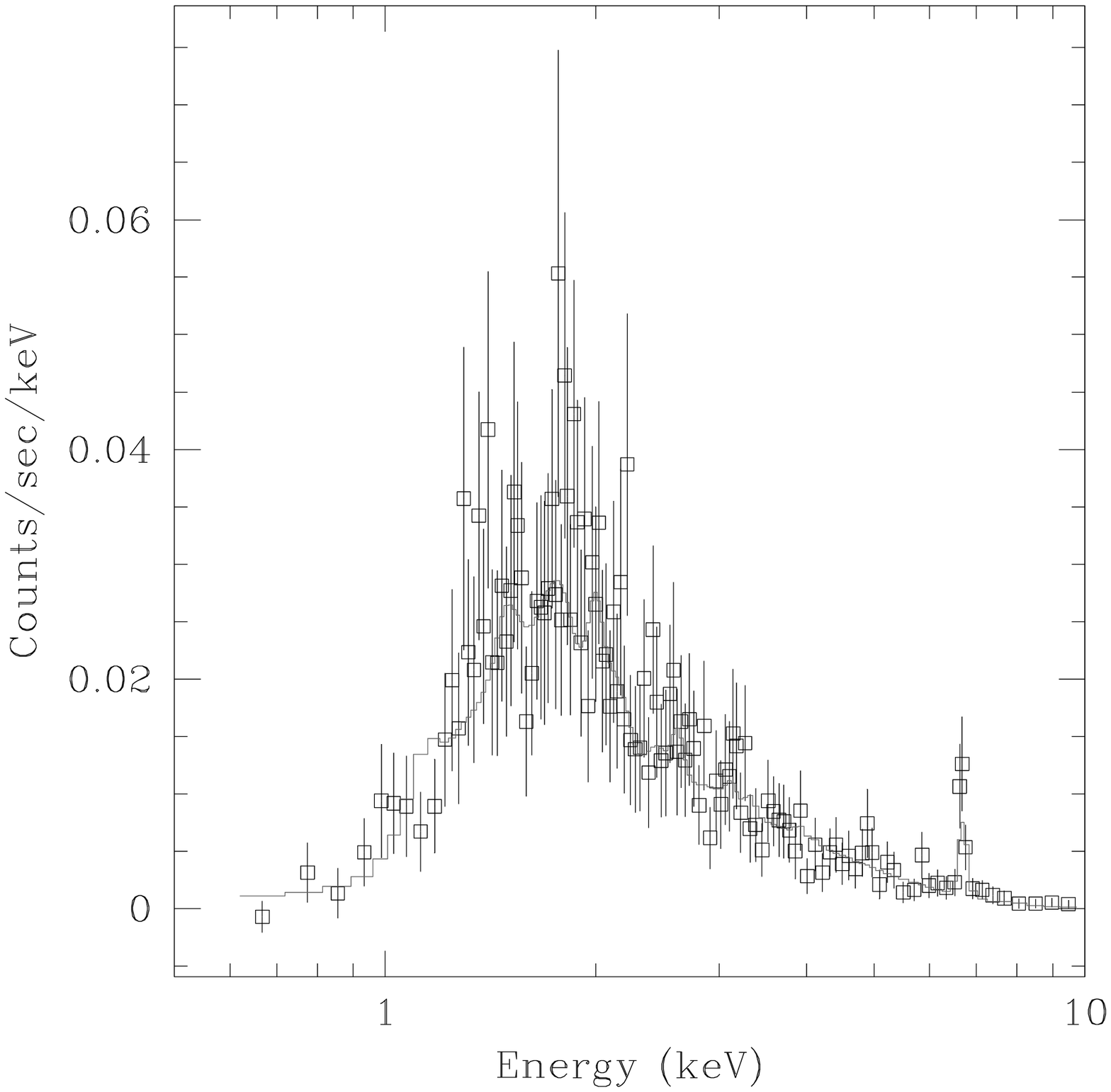}
      }}
      \centerline{X3}
\end{minipage}
\begin{minipage}{3.5cm}
      \centerline{\hbox{
      \includegraphics*[width=3.9cm]{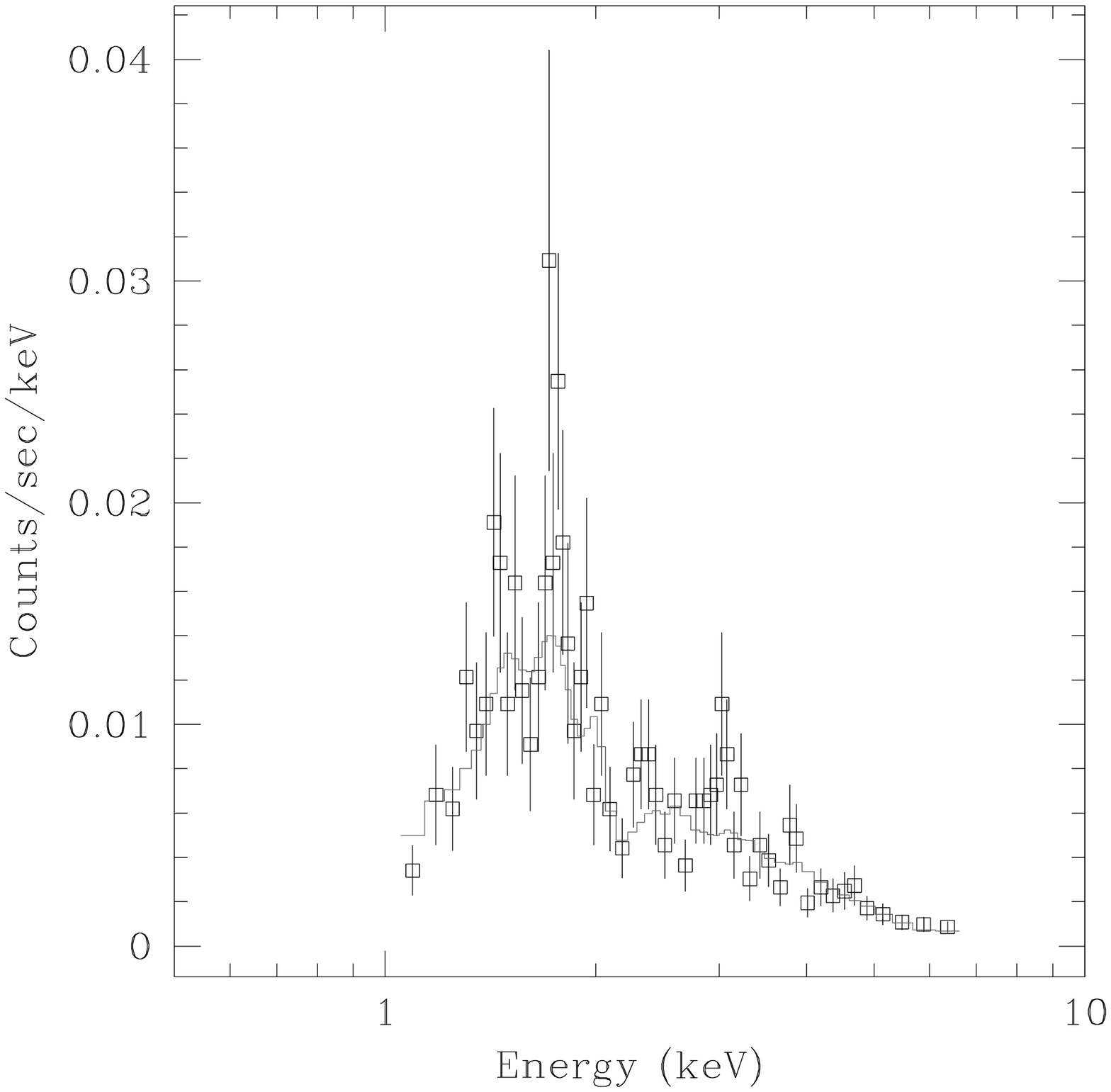}
      }}
      \centerline{C2}
\end{minipage}
%\end{figure}

\vspace{3mm}

%\begin{figure}
\begin{minipage}{3.5cm}
      \centerline{\hbox{
      \includegraphics*[width=3.9cm]{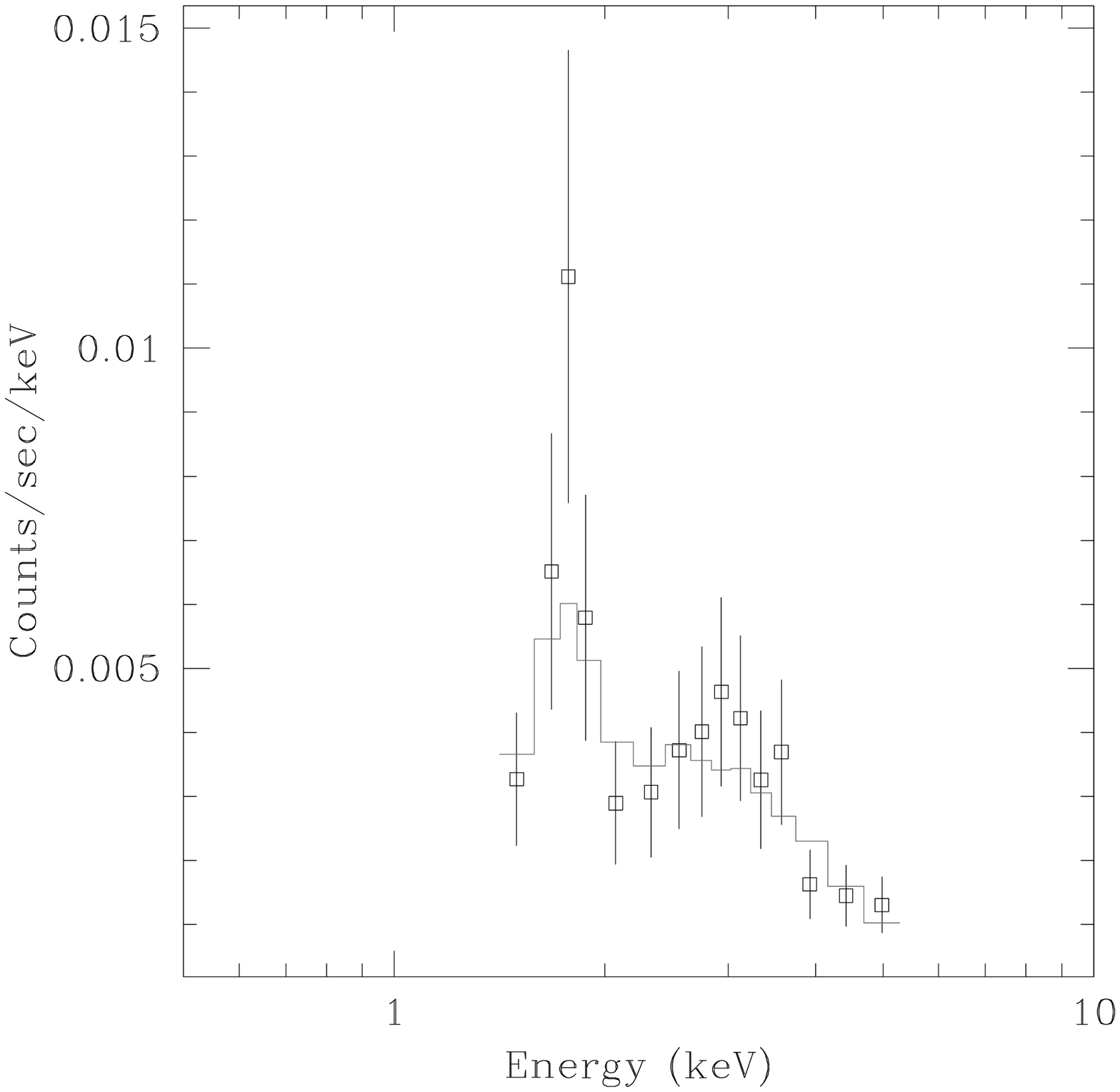}
      }}
      \centerline{IRS 5 : C1}
\end{minipage}
\begin{minipage}{3.5cm}
      \centerline{\hbox{
      \includegraphics*[width=3.9cm]{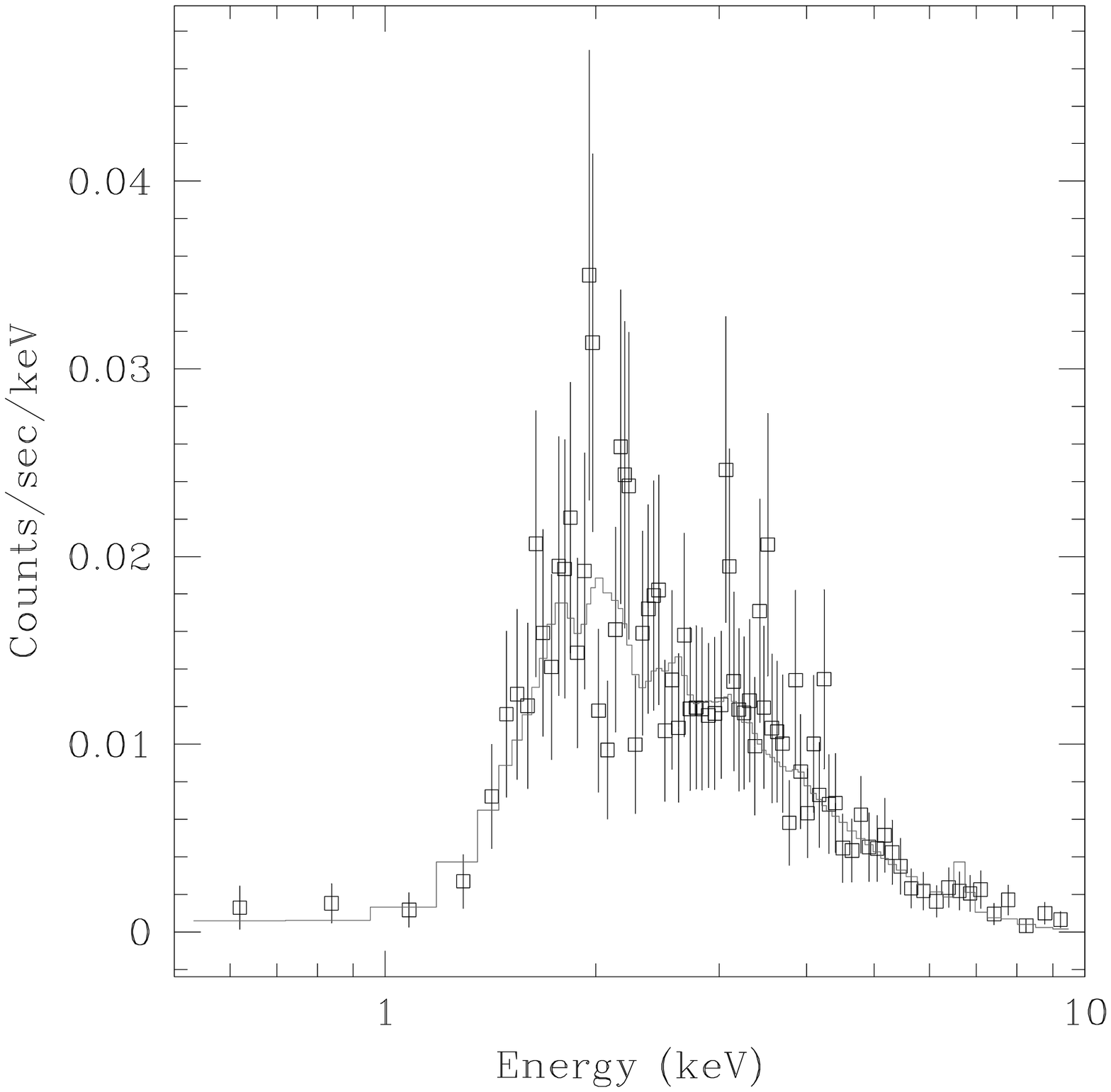}
      }}
      \centerline{X1}
\end{minipage}
\begin{minipage}{3.5cm}
      \centerline{\hbox{
      \includegraphics*[width=3.9cm]{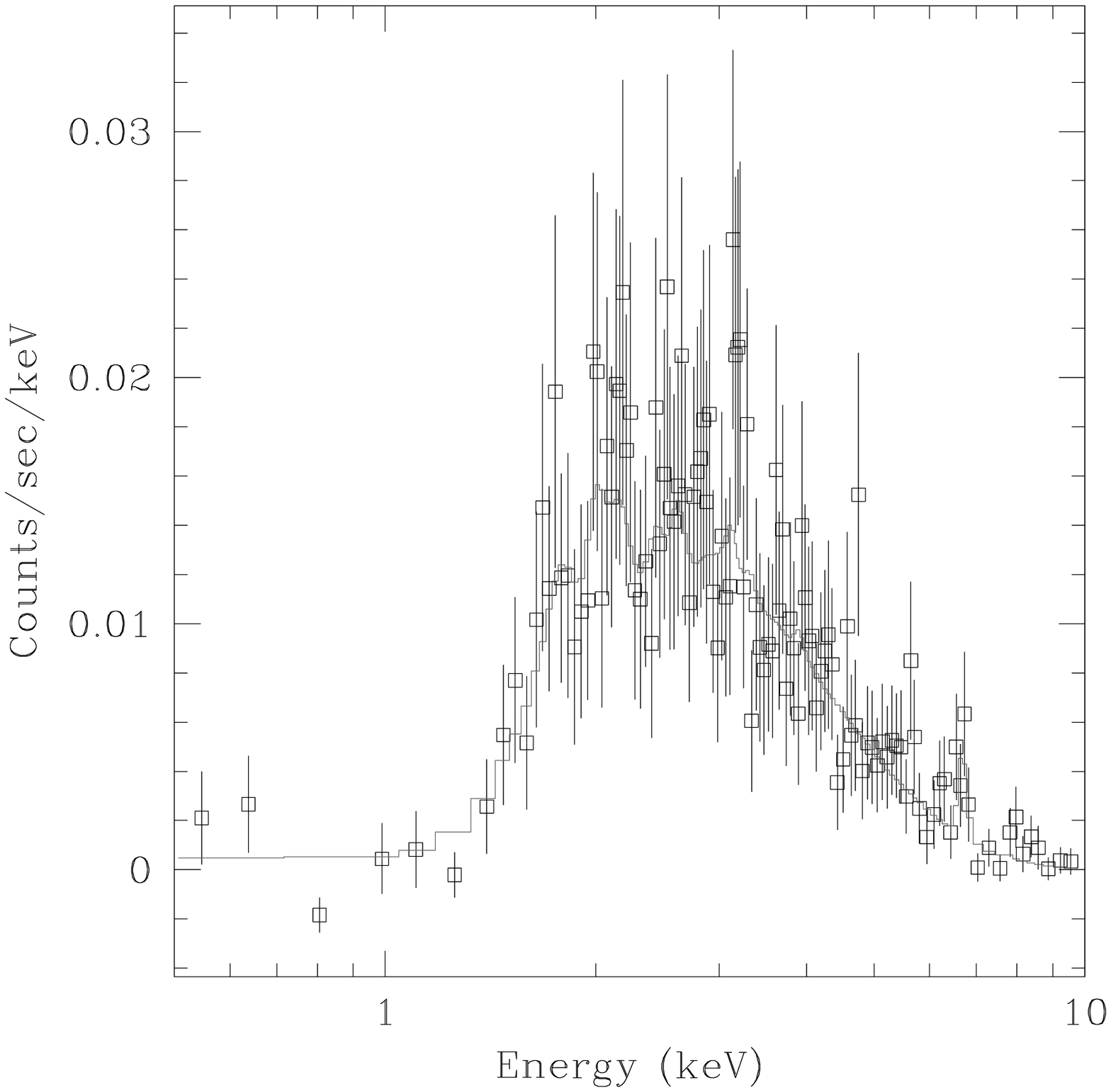}
      }}
      \centerline{X2}
\end{minipage}
\begin{minipage}{3.5cm}
      \centerline{\hbox{
      \includegraphics*[width=3.9cm]{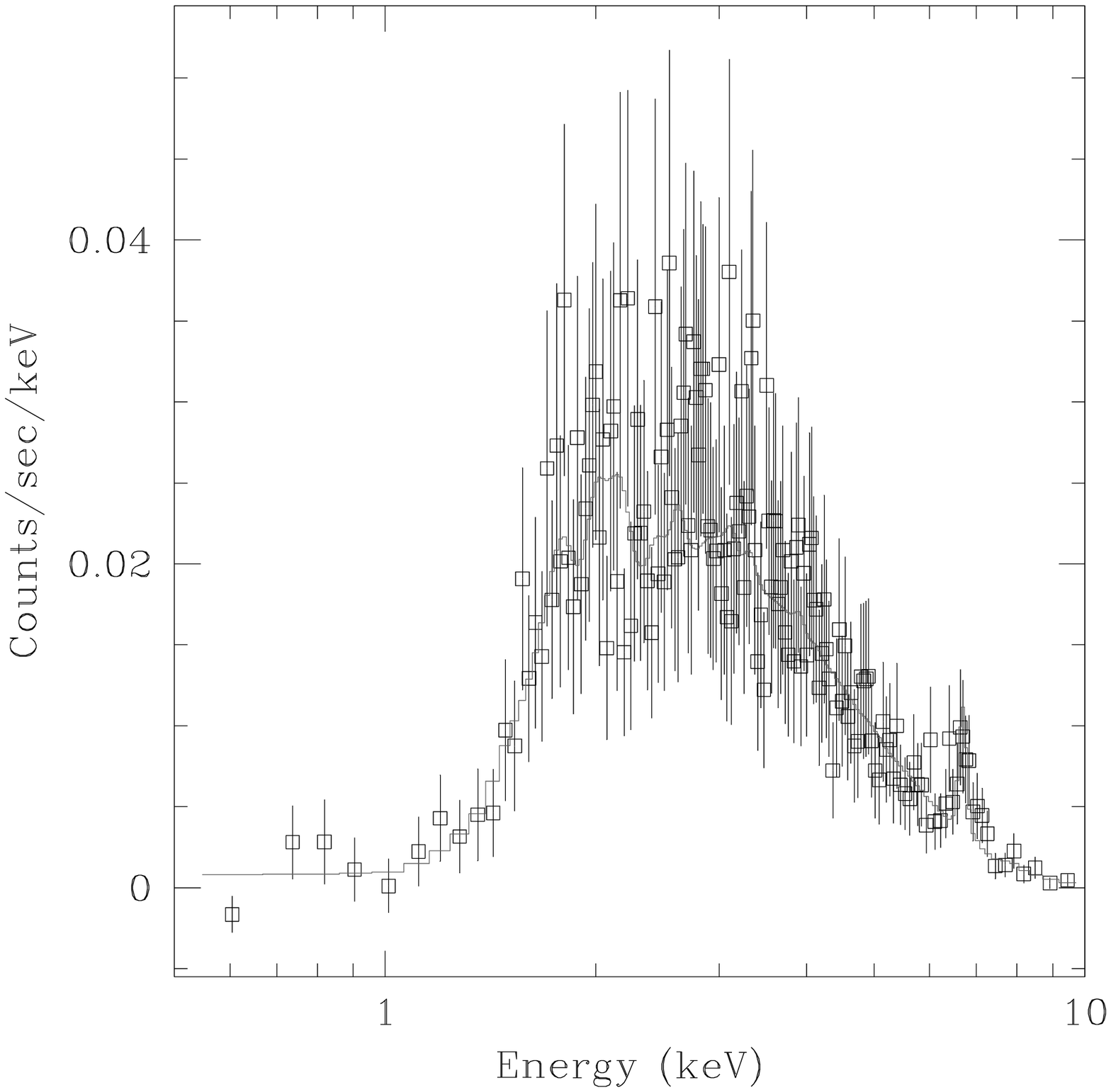}
      }}
      \centerline{X3}
\end{minipage}
\begin{minipage}{3.5cm}
      \centerline{\hbox{
      \includegraphics*[width=3.9cm]{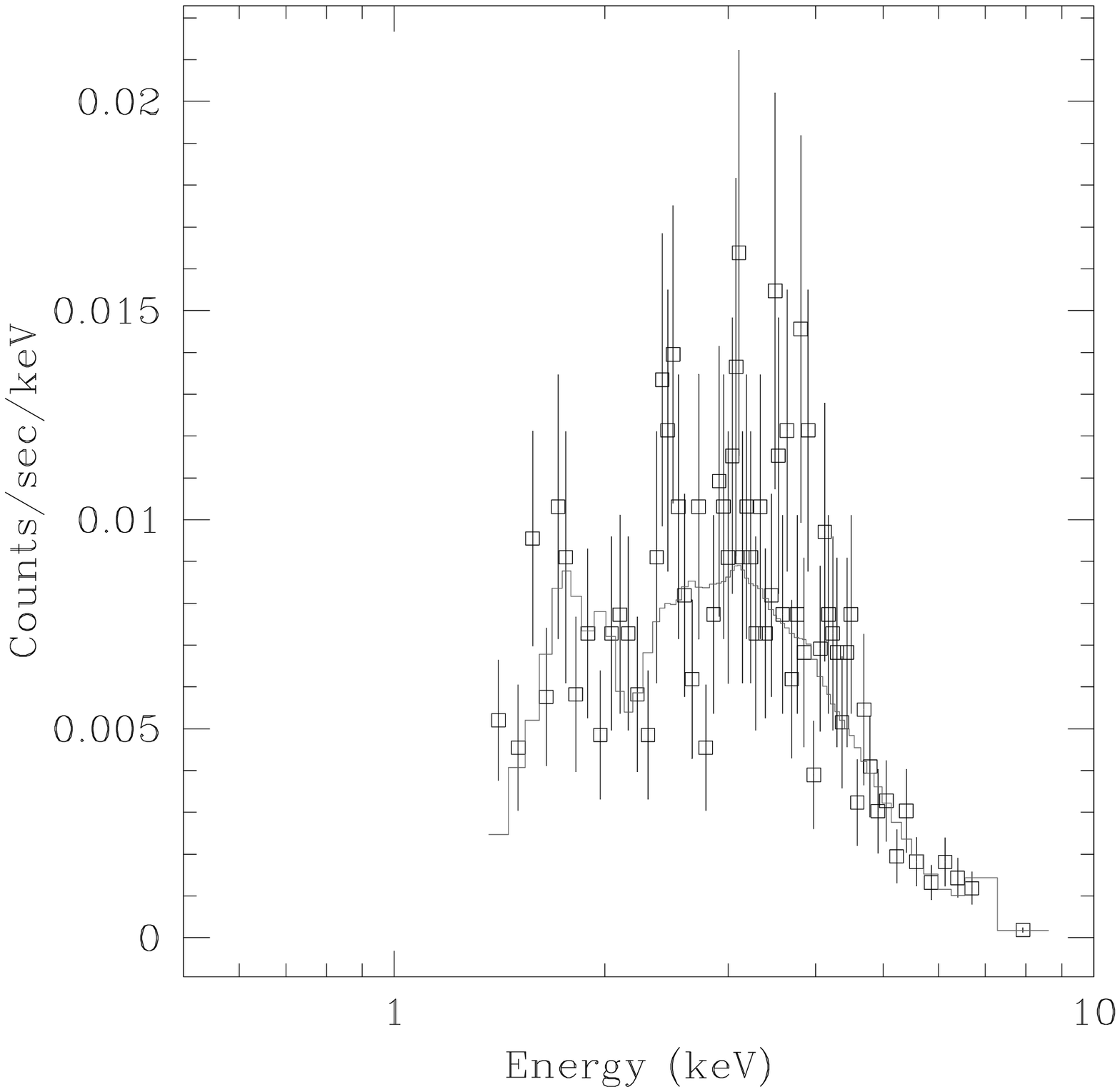}
      }}
      \centerline{C2}
\end{minipage}
%{\bf Fig. 7 FIRST PART - only necessary in referee version}
%\end{figure}

\vspace{3mm}

%\begin{figure}
\begin{minipage}{3.5cm}
      \centerline{\hbox{
      \includegraphics*[width=3.9cm]{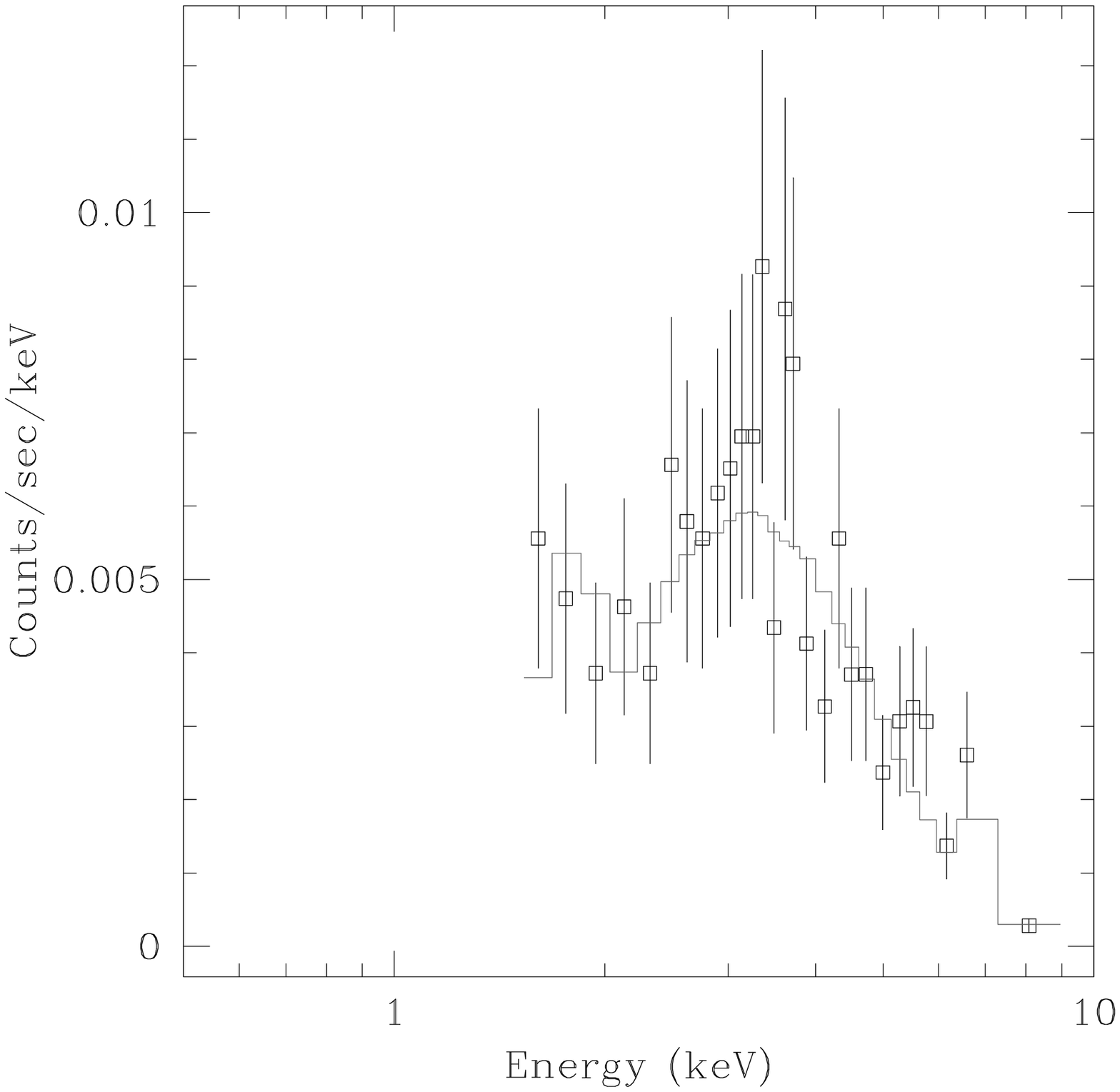}
      }}
      \centerline{IRS 1 : C1}
\end{minipage}
\begin{minipage}{3.5cm}
      \centerline{\hbox{
      \includegraphics*[width=3.9cm]{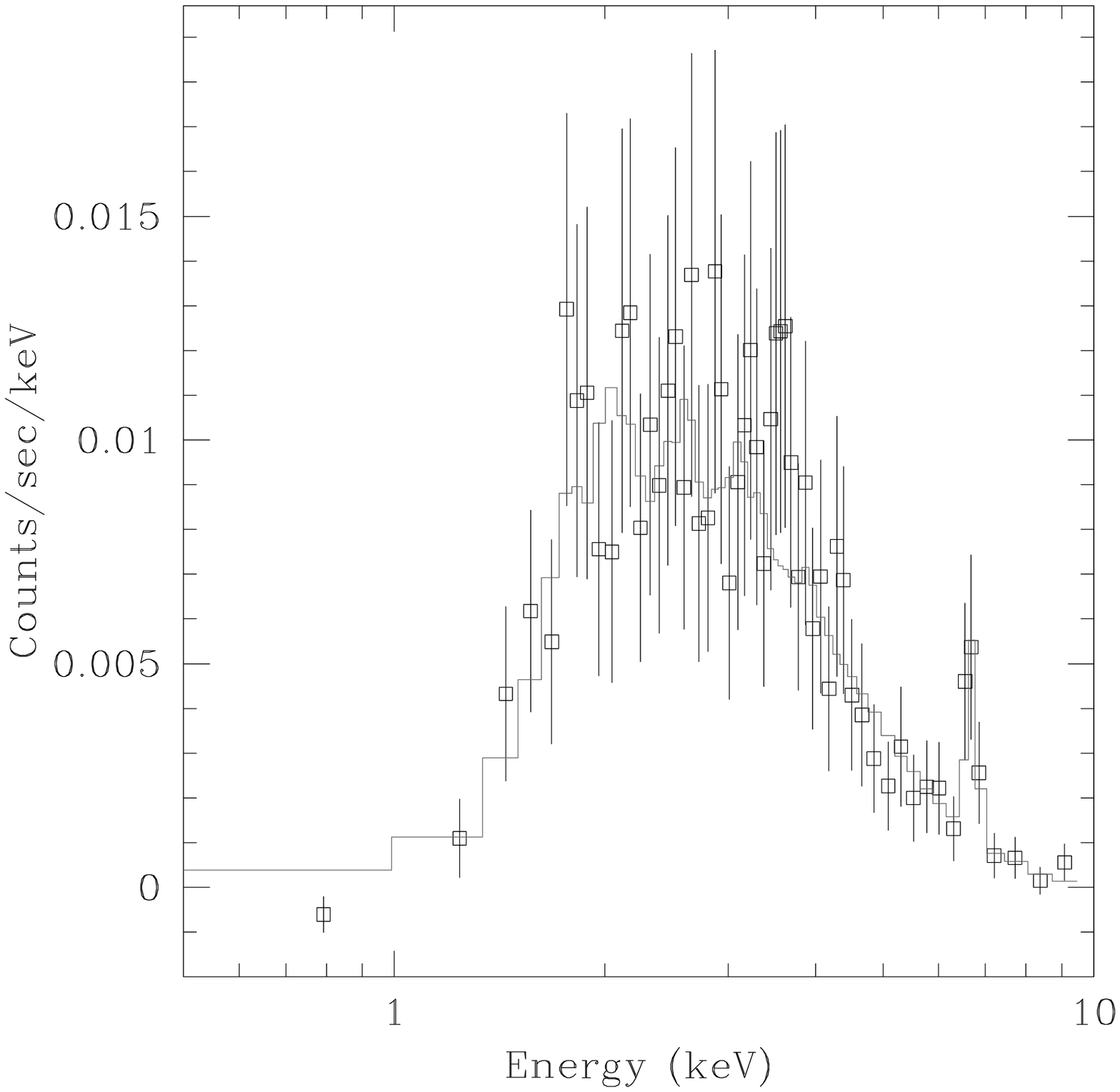}
      }}
      \centerline{X1}
\end{minipage}
\begin{minipage}{3.5cm}
      \centerline{\hbox{
      \includegraphics*[width=3.9cm]{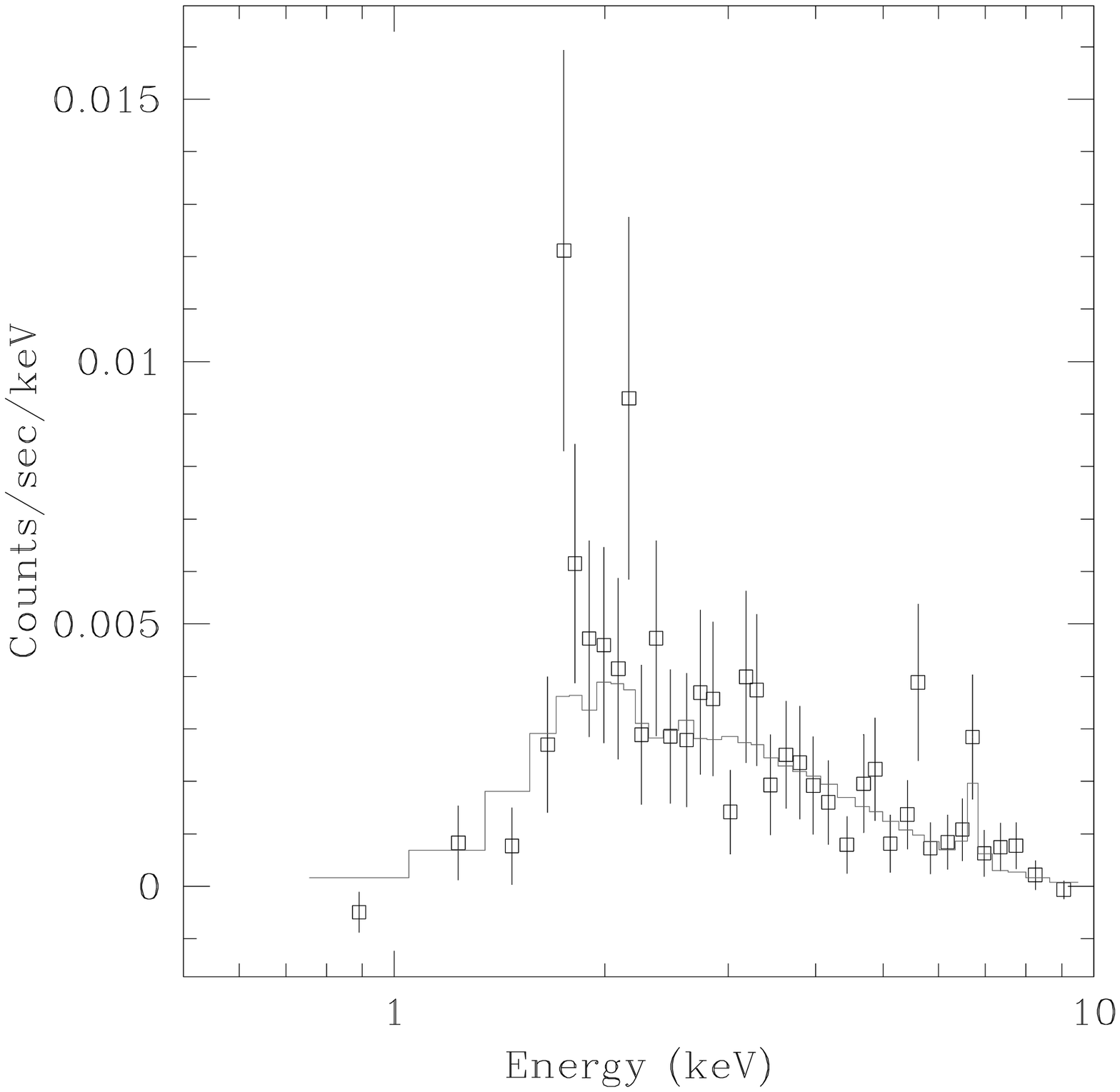}
      }}
      \centerline{X2}
\end{minipage}
\begin{minipage}{3.5cm}
      \centerline{\hbox{
      \includegraphics*[width=3.9cm]{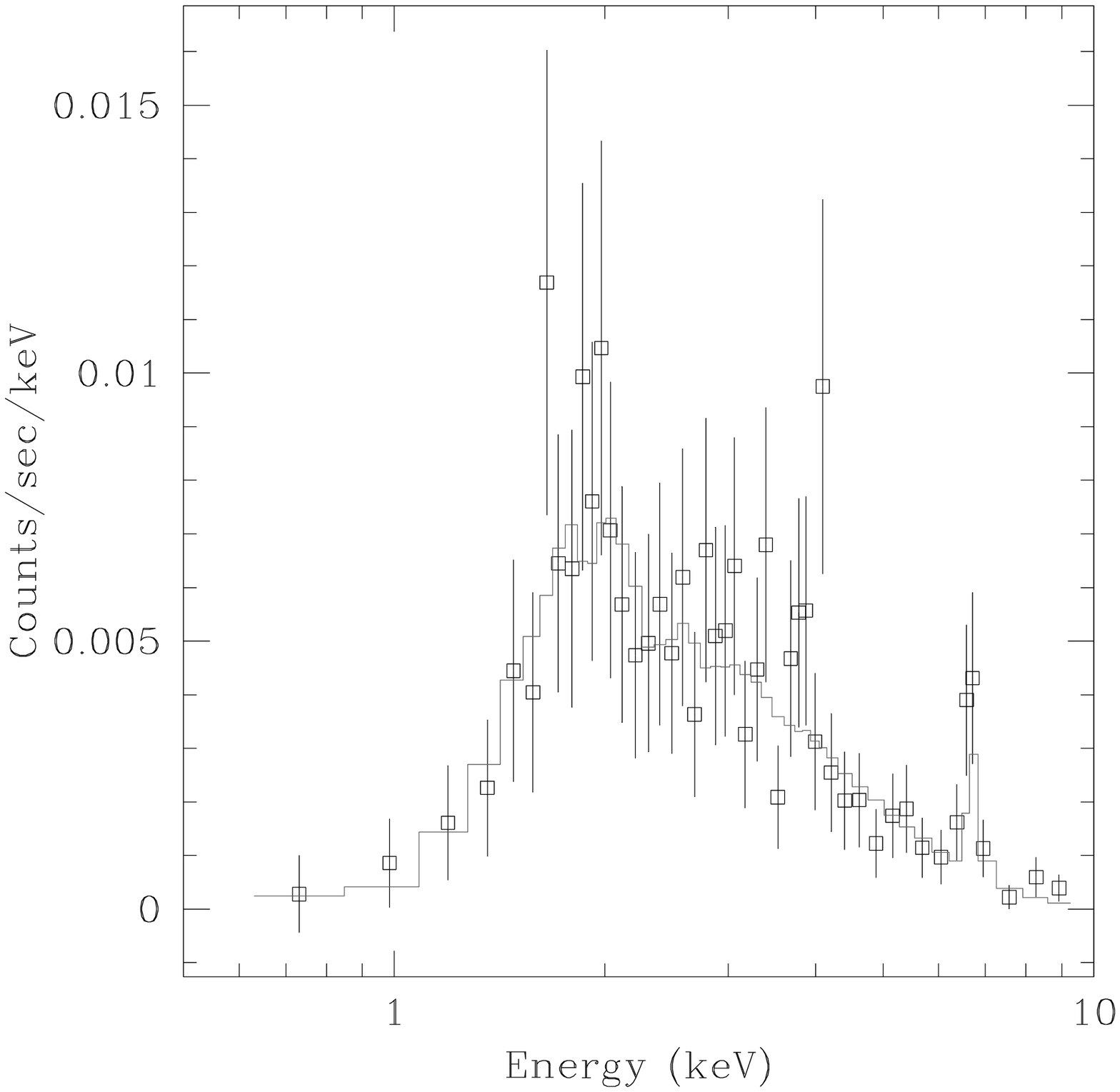}
      }}
      \centerline{X3}
\end{minipage}
\begin{minipage}{3.5cm}
      \centerline{\hbox{
      \includegraphics*[width=3.9cm]{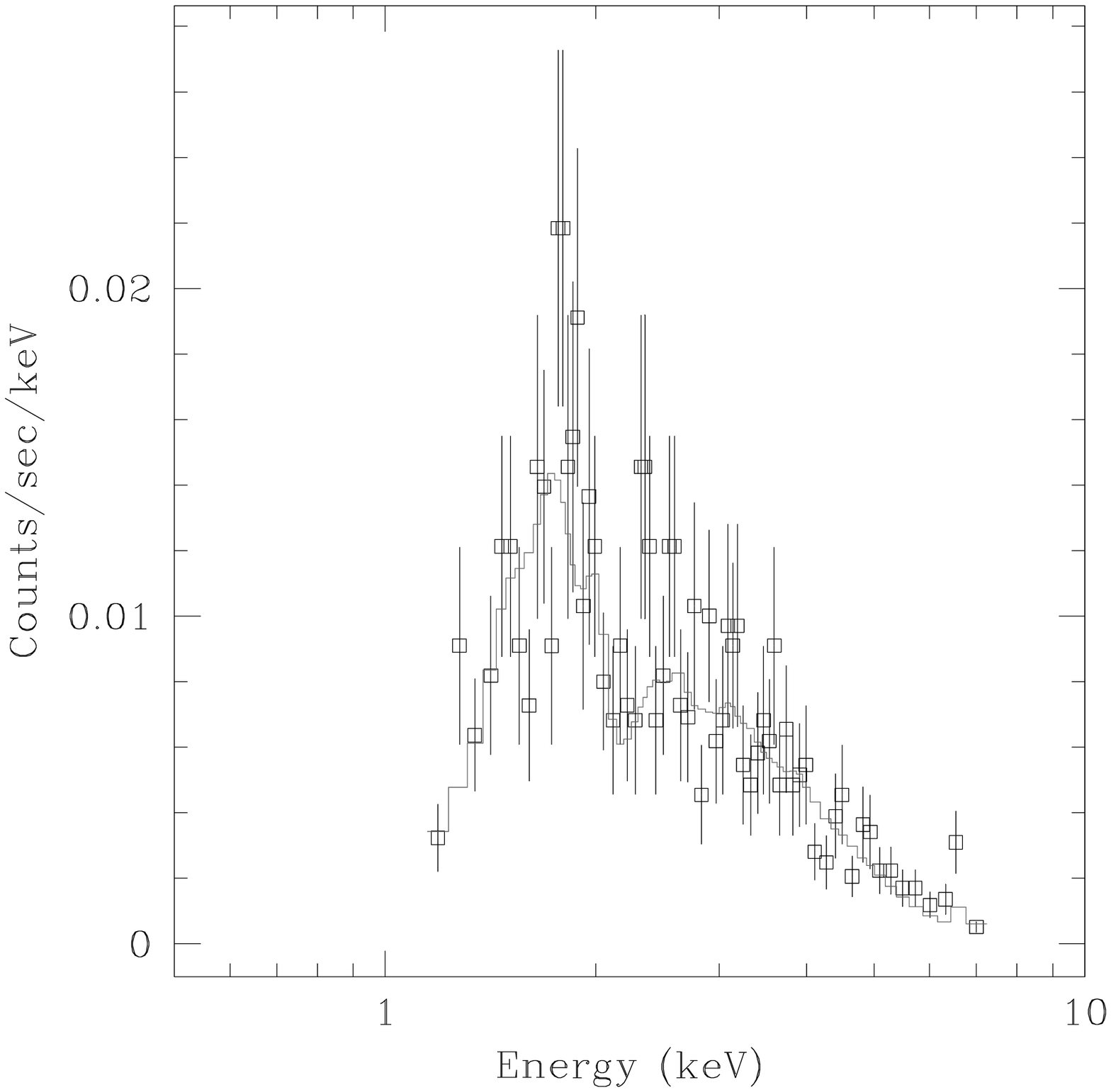}
      }}
      \centerline{C2}
\end{minipage}

%\begin{center}
\caption{X-ray spectra of IRS~2, 5 and 1 derived from the five X-ray
datasets. The line shows the result of fitting an absorbed APEC model.}
\label{xrayspectra}
%\end{center}
\end{figure*}

\begin{figure*}
% \begin{center}
\centering
  \includegraphics[width=16cm]{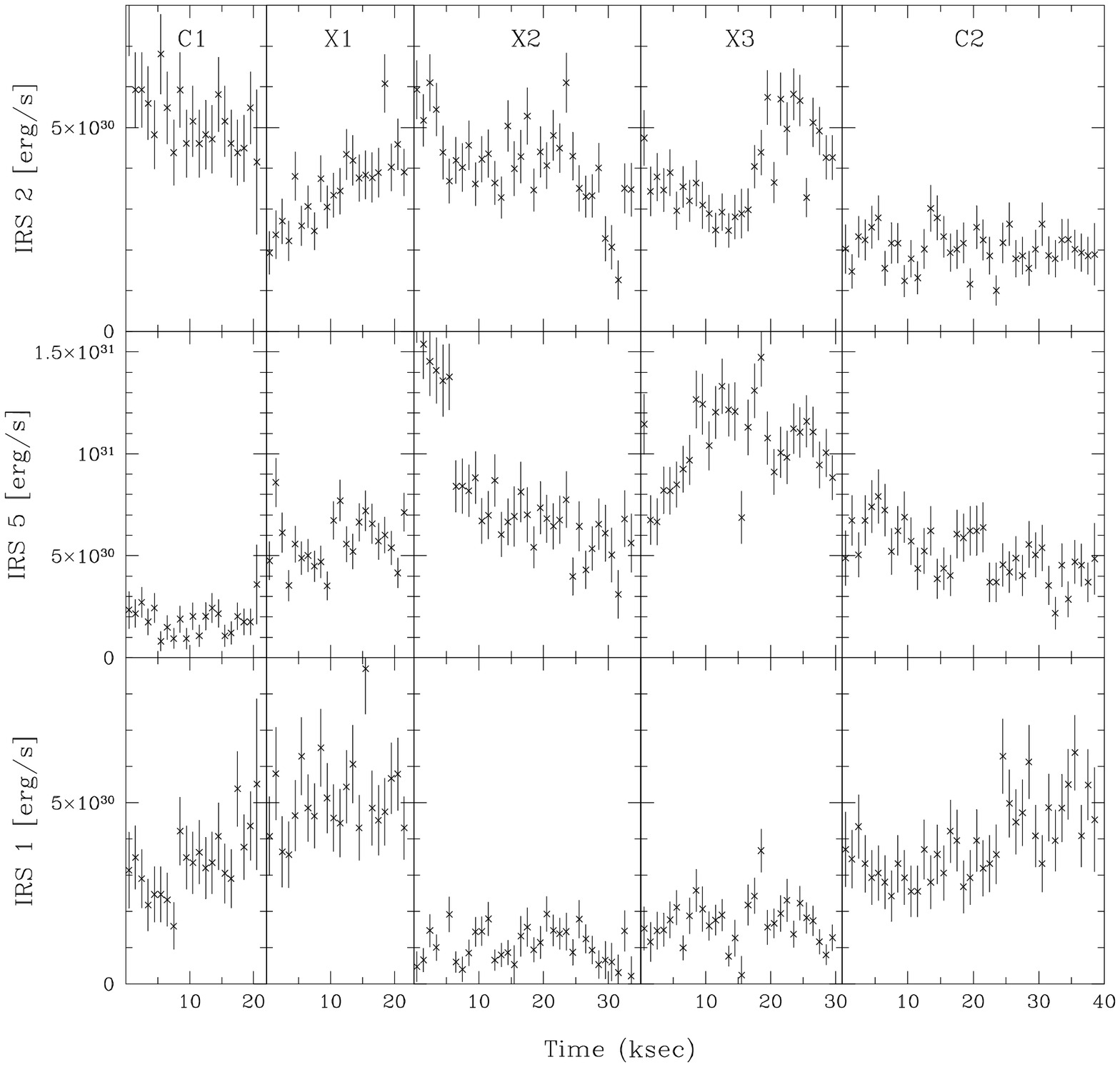}
  \caption{Kilosecond-binned, background-subtracted X-ray luminosity curves 
  in temporal order for the three class~I protostars IRS~2,5, and 1. 
  In the X2 and X3 observations, around 20\% of the flux from IRS~1 is
  lost due to a chip gap in the source region.}
\label{lcslumi}
% \end{center}
\end{figure*}

At X-ray wavelengths, nine sources are clearly detected, including an unidentified
weak source (no. 3) only detected by \textsl{Chandra} (see Tables~\ref{sourceidlist} 
and \ref{sourcexrlist}). Fig.~\ref{chandraview} shows the combined \textsl{Chandra} 
data, with the combined XMM-\textsl{Newton} data shown as an inset. It is interesting
to compare this X-ray view with the radio view in Fig.~\ref{vlainset}. The five
weak radio sources only emerging in the complete VLA data all remain undetected
in X-rays together with the Radio Sources 5 and 9 \citep{bro87}. However, the
class~I protostar IRS~9 -- clearly detected in X-rays, even with a flare in the
C1 observation -- remains undetected at radio wavelengths, together with the weak
X-ray source no.3 not showing any counterpart at other wavelengths. 
Comparing the count rates for these sources derived from the five X-ray datasets
shows that there is considerable variability in many cases (Table~\ref{sourcexrlist}). 
For the three class~I protostars IRS~1, 2, and 5, this is corroborated by the
luminosity curves (Fig.~\ref{lcslumi}) and the corresponding spectra
(Fig.~\ref{xrayspectra}). All three sources have X-ray luminosities
on the order of several $10^{30}$~erg/s. IRS~5 emerges again as the most
variable source.

\begin{figure*}
\begin{minipage}{6cm}
      \centerline{\hbox{
      \includegraphics*[width=7cm]{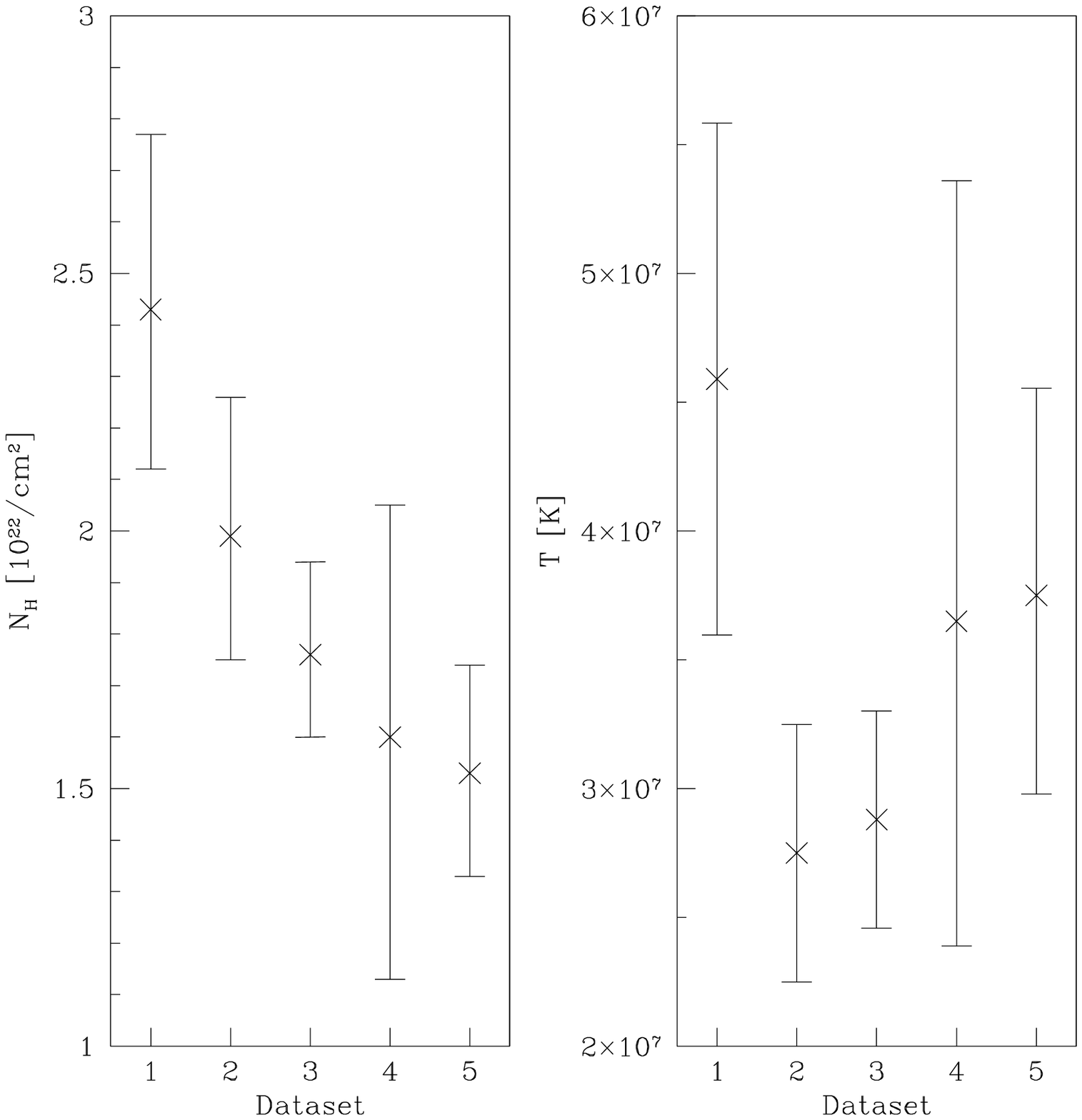}
      }}
      \centerline{IRS~2}
\end{minipage}\    \
\begin{minipage}{6cm}
      \centerline{\hbox{
      \includegraphics*[width=7cm]{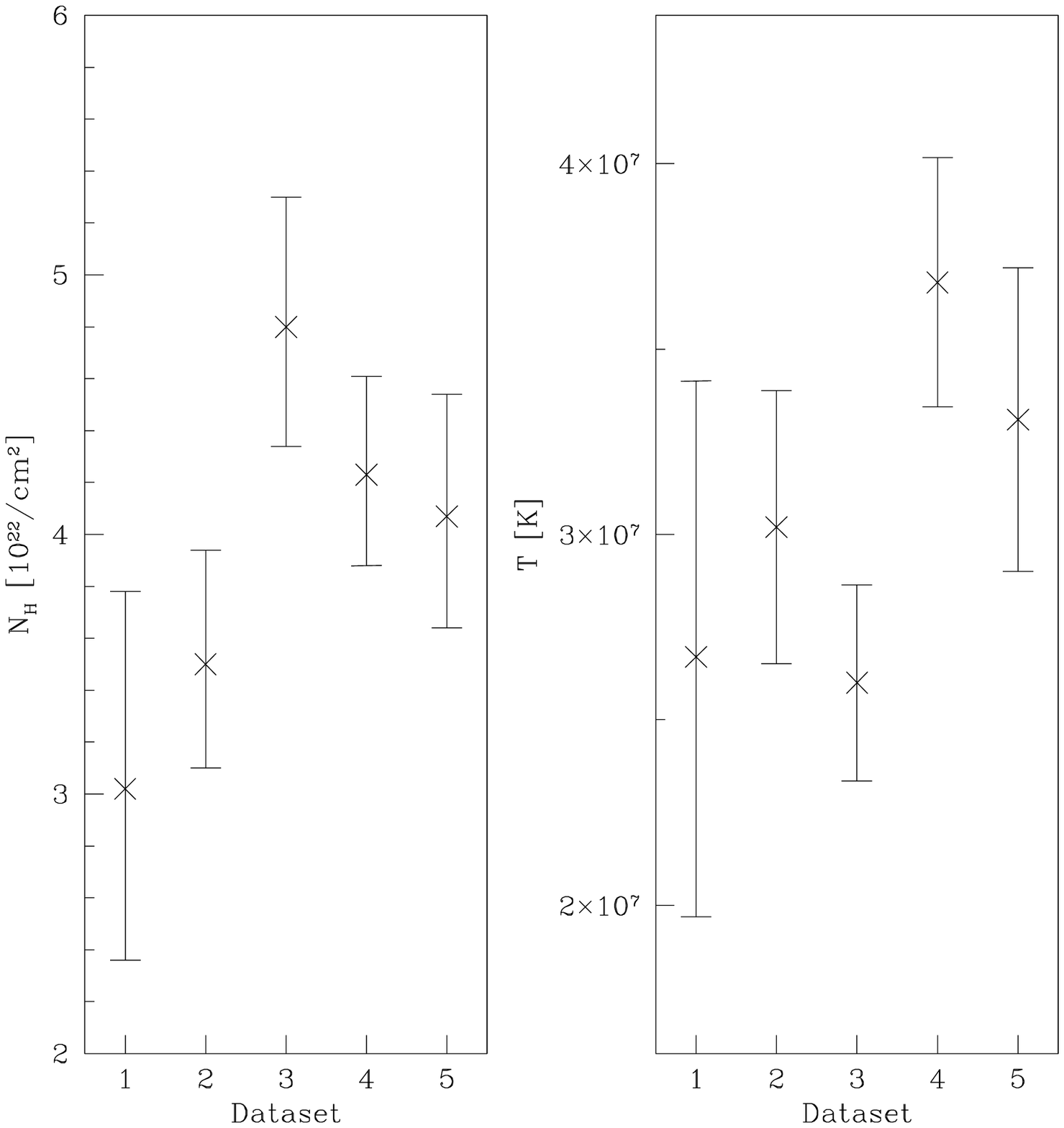}
      }}
      \centerline{IRS~5}
\end{minipage}
\begin{minipage}{6cm}
      \centerline{\hbox{
      \includegraphics*[width=7cm]{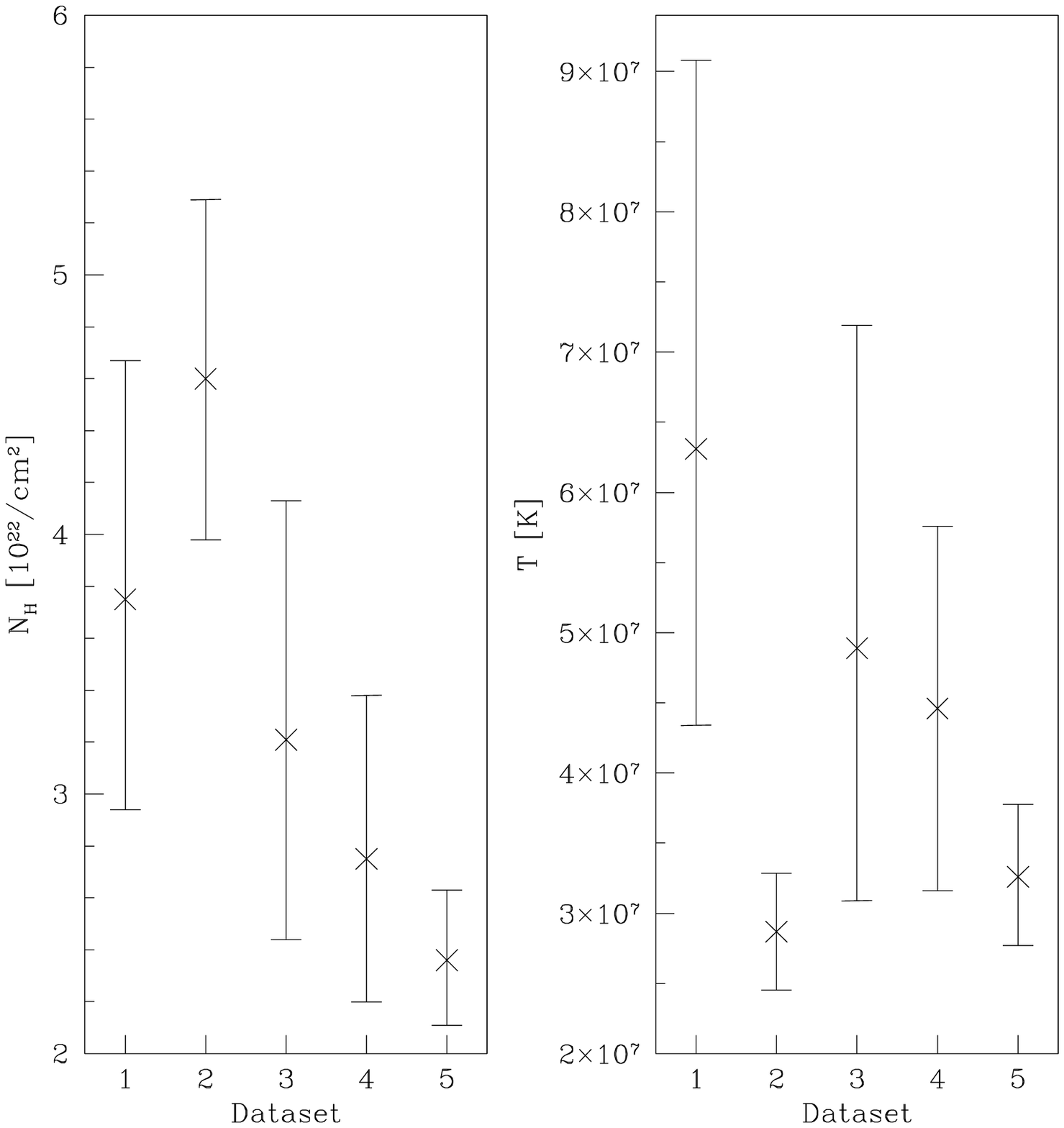}
      }}
      \centerline{IRS~1}
\end{minipage}

%\begin{center}
\caption{Results of fitting XSWABS$\times$XSAPEC, fit parameters
N$_{\rm H}$ and kT. Abundances set to values determined from best
spectra. Errorbars are 3$\sigma$. The X-ray datasets are numbered
according to Table~\ref{xraylist}. The NIR-determined N$_{\rm H}$
column densities from \citet{nis05} are $6 \times
10^{22}$~cm$^{-2}$ (IRS~1), $4 \times 10^{22}$~cm$^{-2}$ (IRS~2) and
$9 \times 10^{22}$~cm$^{-2}$ (IRS~5).}
\label{xrayfitpars_v}
%\end{center}
\end{figure*}

The elemental abundances for the three sources, as derived from the
6.7~keV Fe~K line emission, are $0.5 \pm 0.2$ (IRS1,X1), $0.7 \pm 0.1$
(IRS2,X3) and $0.26 \pm 0.05$ (IRS5,X3) relative to solar abundances.
There is no evidence for fluorescent emission of weakly ionized or
neutral iron at energies below 6.7~keV. Fitting absorbed APEC
emission models to the spectra leads to the results depicted in
Fig.~\ref{xrayfitpars_v} and listed in Table~\ref{fitrestab}. All three
sources show high absorbing column densities (several $10^{22}$~cm$^{-2}$) 
and hot plasma emission (several $10^7$~K). There are signs of long-term
variability in the absorbing column densities, especially towards IRS~2
and IRS~1. Note that the X2 and X3 observations are consecutive while a
total range of about 2.5 years is covered. Tentatively, there is a trend 
for the emission being harder when the sources are brighter, although the
situation is inconclusive.

Interestingly, though, the hydrogen column densities as derived from the 
X-ray spectra for IRS~1,2, and 5 are at around half the values derived by \citet{nis05}
from near-infrared colours\footnote{adopting ${\rm A_V}=15.9 \cdot E(H-K)$ \citep{rie85} } (see Table~\ref{compareNH}). 
In order to check the significance of this discrepancy, we tried to fit the spectra while freezing the absorbing column density $N_H$ to the value derived from NIR colours. This lead to very poor fits (large $\chi^2$) in all cases, demonstrating that the large $N_H$ values are inconsistent with the X-ray spectra.
Additionally, by using a Monte Carlo fitting scheme, the parameter space was checked for multiple minima, but again no acceptable model was found for large $N_H$ values. 
The empirical relation is 
$N_H[{\rm cm^{-2}}] \approx 2 \times 10^{21} \times A_V[{\rm mag}]$
\citep{ryt96,vuo03}. A similar discrepancy has been observed
previously towards the young stellar objects L1551 IRS~5 \citep{bal03}, 
EC~95 \citep{pre03b}, and towards SVS~16 \citep{pre03a}. Possible solutions
to this 'extinction problem' are configurations in which the protostar and its
X-ray emission become detached, so that the absorbing column densities for the
infrared and X-ray emission become different. Scenarios include X-ray 
emission from jet shocks close to the protostar, X-rays scattered towards
the observer by circumstellar material, and huge coronal structures.

\begin{table}
%\begin{center}
\caption[]{Comparison of column densities towards IRS 2, 5, and 1 derived from
NIR and X-ray emission}
%/ XXX Dust: $T_d= 20$~K, opacities O\&H94, MH/Mdust=110}
\begin{tabular}{lrr}
\hline
Source & $N_H$(NIR)$^{\rm a}$& $N_H$(X-ray)   \\% & $N_H$(dust)    \\
       & [10$^{22}$~cm$^{-2}$]     & [10$^{22}$~cm$^{-2}$] \\%& [10$^{22}$~cm$^{-2}$]\\
\hline
\hline
IRS 2  & $4.0 \pm 0.6$         & $1.9 \pm 0.4$         \\%& $26.0$ \\
IRS 5  & $9.0 \pm 0.6$         & $3.9 \pm 0.7$         \\%& $10.4$ \\
IRS 1  & $6.0 \pm 0.6$         & $3.3 \pm 0.9$         \\%& $19.5$ \\
%extinctprobvgl.sxc
\hline
\label{compareNH}
\end{tabular}

$^{\rm a}$ $A_V$ (from \citealp{nis05}) converted into $N_H$ \\
using $N_H[{\rm cm^{-2}}] \approx 2 \times 10^{21} \times A_V[{\rm mag}]$, see text

%\end{center}
\end{table}
%NEW

\begin{figure}
% \centering
 \includegraphics*[width=9cm]{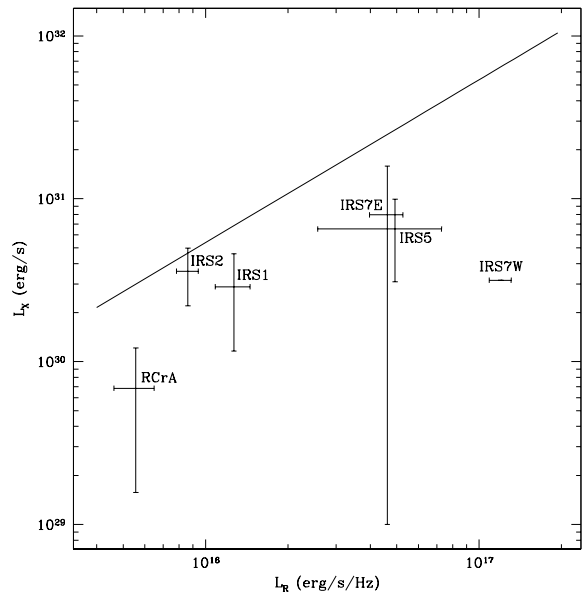}
% \resizebox{\hsize}{!}{\includegraphics*[width=9cm]{feigel98et.eps}}
 \caption{X-ray vs. radio luminosity of identified young stellar objects
for which X-ray luminosities can be derived from spectra. The errorbars correspond 
to the standard deviation within the 1998 VLA and archival X-ray data, except for 
IRS~7E/W, where the X-ray luminosities are from \citet{ham05}. For IRS~7E, the X-ray
errorbars reflect the flare seen in X2/X3 while for IRS~7W, only a single luminosity
was determined from the X2 data. The line shows the empirical relationship for 
active late-type stars from \citet{beg94}.}
\label{plotLXLR}
\end{figure}

For six of the young stellar objects studied here, it is possible to reliably 
determine X-ray luminosities based on their spectra. The radio and X-ray luminosities
of these sources have been plotted into Fig.~\ref{plotLXLR} together with an empirical 
relationship for active late-type stars from \citet{beg94}. The underlying physics
of this relation is that the acceleration of synchrotron-emitting electrons and the
heating of the coronal plasma leading to X-ray emission are accomplished by the same
mechanisms. The \textsl{Coronet} YSOs appear to be compatible with this 
relationship.

Comparing the X-ray luminosities to the bolometric luminosities (see Table~\ref{lumitab}), it turns out that IRS~2, 5, 1, and R CrA are all in the unsaturated regime $L_{\rm X}/L_{\rm bol}<10^{-3}$ 
\citep{fle95}, but a lot more active than the sun 
($(L_{\rm X}/L_{\rm bol})_{\odot} \approx 10^{-6}$), IRS~5 
being the most active source with a ratio of log($L_{\rm X}/L_{\rm bol})=-3.4$.

\begin{table}
%\begin{center}
\caption[]{Bolometric and X-ray luminosities (derived from X-ray spectra)}
\begin{tabular}{lrrr}
\hline
Source & $L_{\rm bol}^{\rm a}$ & $L_{\rm X}$ & log($L_{\rm X}/L_{\rm bol})$\\
       & L$_\odot$     & $10^{-3}$~L$_\odot$ \\
\hline
\hline
IRS 2 & 16 & 0.93 & $-4.2$\\
IRS 5 &  4 & 1.70 & $-3.4$\\
IRS 1 & 19 & 0.75 & $-4.4$\\
R CrA & 132& 0.18 & $-5.9$\\
%RCRA: Lorenzetti+99, Rest Wil92 skaliert
%corolumis.sxc
\hline

\label{lumitab}
\end{tabular}

$^{\rm a}$ scaled to $d=150$~pc from \citet{wil92}, \\
for R~CrA directly from \citet{lor99}
%\end{center}
\end{table}
%NEW

\subsection{Synthesis}

\begin{figure*}
\begin{minipage}{9cm}
      \centerline{\hbox{
      \includegraphics*[width=9cm]{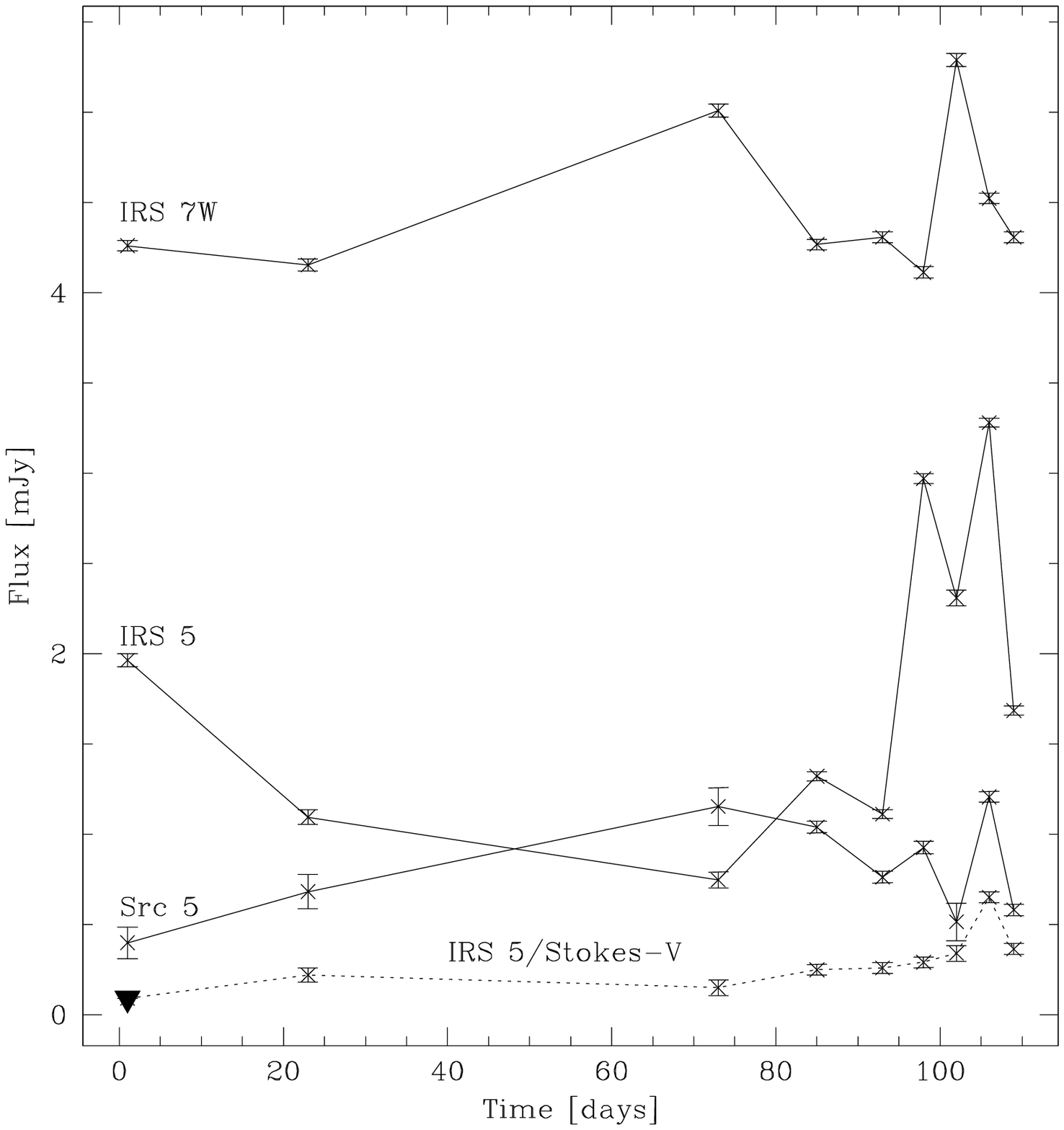}
      }}
      %\centerline{Text}
\end{minipage}\    \
\begin{minipage}{9cm}
      \centerline{\hbox{
      \includegraphics*[width=9cm]{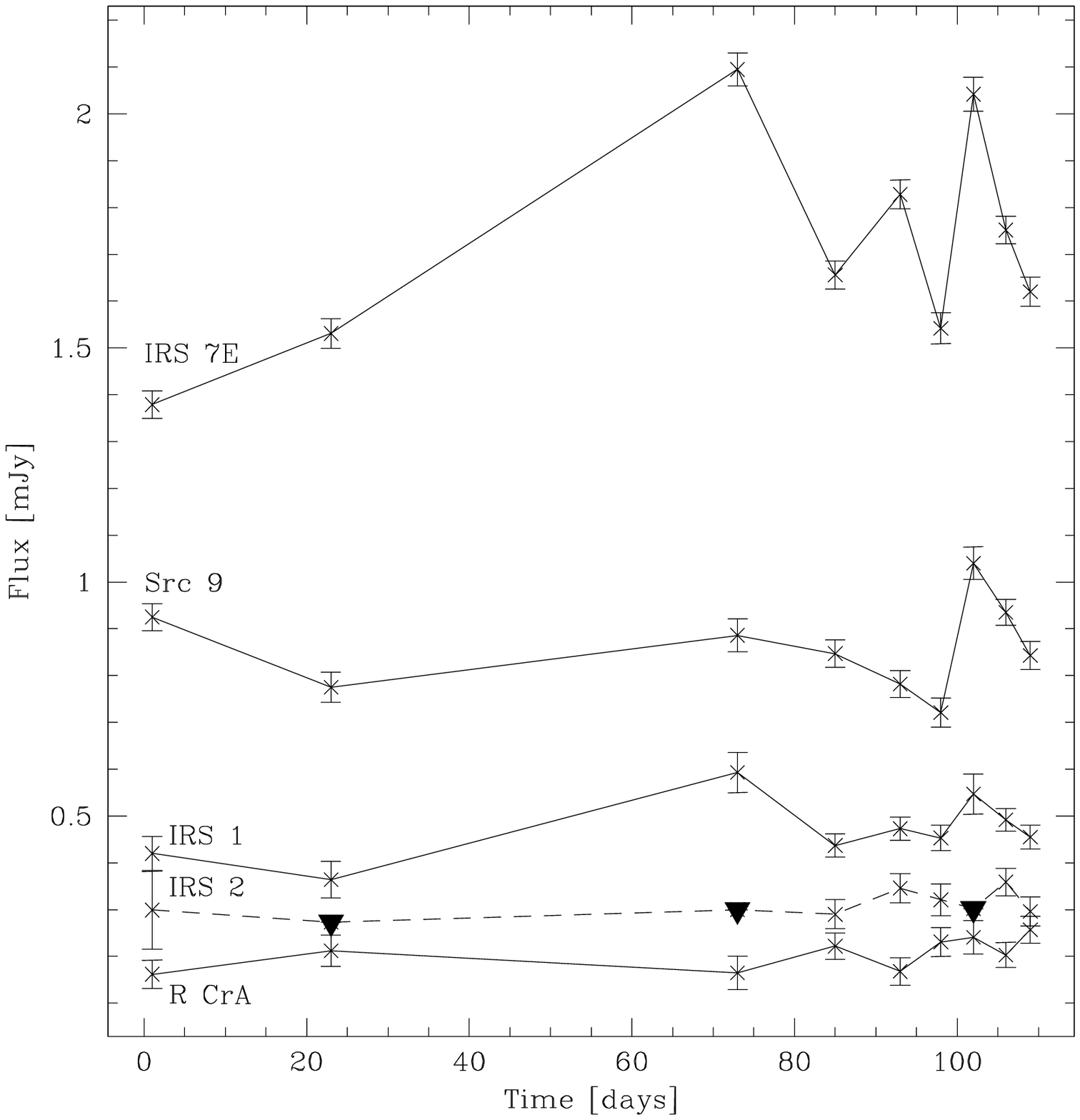}
      }}
      %\centerline{Text}
\end{minipage}
%\begin{center}
\caption{Radio flux curves of sources identified in the 1998
VLA data. Error bars indicate the uncertainty estimates
given by the AIPS task 'sad', they are based on actual noise ($1\sigma$) rather than
the quality of the fit. Filled triangles denote upper limits.
Also shown is Stokes-$V$ for IRS~5.}
\label{vlacurves}
%\end{center}
\end{figure*}

\begin{figure}
 \includegraphics*[width=9cm]{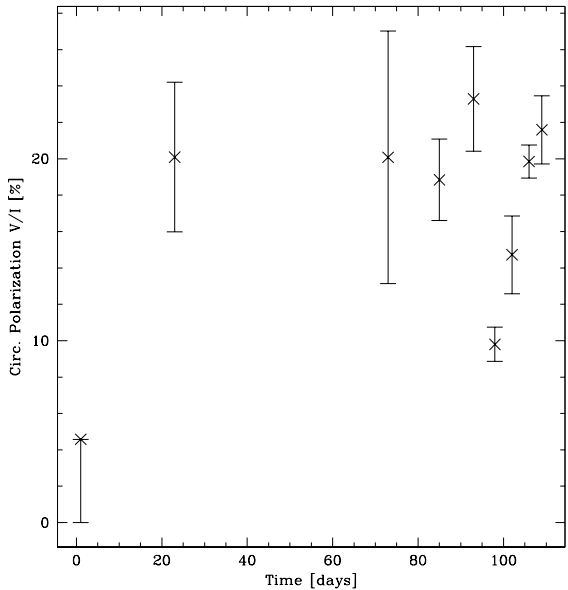}
 \caption{IRS~5 epoch-averaged polarization degrees in percent. Errors are propagated from those shown in Fig.~\ref{vlacurves}.}
\label{irs5pol}
\end{figure}

\subsubsection{The class I protostars IRS~2, IRS~5, IRS~1, and IRS~9}

IRS 2 does not show significant radio
variability in any time interval covered by our observations (Fig.~\ref{vlacurves} and Table~\ref{sourceralist}). The flux
is, however, at about half the value determined by \citet{fcw98}. 
IRS~2 is one of the strongest X-ray sources among the objects studied here. 
The overall count rate in the C1 \textsl{Chandra} observation is
twice as high as during the C2 observation which is clearly visible in the 
luminosity curves (Fig.~\ref{lcslumi}). The XMM-\textsl{Newton} count rates 
are less diverse. Variability is seen also within observations, especially X1, 
showing a continuous increase, and X3, showing a sudden increase in the second 
half of the observation. Already the ROSAT data presented by \citet{nep97} 
show variability on timescales of months.
The X-ray spectra (Fig.~\ref{xrayspectra}), can be explained as the emission of hot plasma ($T \approx 4 \times 10^7$~K), absorbed by a column density of $N_H \approx 2 \times 10^{22}$~cm$^{-2}$. Apparently, the absorbing column density towards IRS~2 has continuously diminished over two and a half years, the 3$\sigma$ errorbars of the first and the last measurements are clearly separated (Fig.~\ref{xrayfitpars_v} and Table~\ref{fitrestab}). This could be caused by a clumpy circumstellar medium, probably involving the circumstellar disk. \citet{nis05} find that accretion accounts for 60\%
of this object's bolometric luminosity.

\vspace*{3mm}
\noindent
IRS~5 shows several flares in our VLA
dataset (Figs.~\ref{vlacurves} and \ref{irs5pol}, Table~\ref{sourceralist}) and has been observed flaring already once before at 4.9~GHz
\citep{sut96}. In our data, the source starts at a relatively high flux, dims, and 
then sharply rises towards two peaks, separated by about nine days.
This source was previously reported to emit nonthermal radio emission \citep{fcw98}. 
Interestingly, the initial high flux of IRS~5 is accompanied by undetectable
Stokes-$V$ emission. Also when looking at the double peak towards the end of the
period covered here, the Stokes-$V$ flux appears to be largely uncorrelated to the
Stokes-$I$ flux: Only in the second peak there is a rise in Stokes-$V$ flux.
Possibly this different behaviour is at least partly due to an averaging effect,
given the faster variability in Stokes-$V$: Contrary
to the Stokes-$I$ emission, Stokes-$V$ in our data is variable down to
timescales of 0.5h, a clear sign of nonthermal emission. 
In the epoch with the highest Stokes-$V$ flux (R8), for example, Stokes-$V$ emission
was only detected in the first quarter. Two epochs earlier, at the sharp rise in Stokes-$I$,
the polarization degree drops from 23\% to 10\% because only the Stokes-$I$ emission increases.
In the subsequent epoch, the polarization degree is back at 20\% while the total
flux increased even further. \citet{fcw98} reported a polarization degree
changing between 10\% to 37\% during a day.

In X-rays, IRS~5 is the most variable object when compared to IRS~1
and IRS~2, within our datasets as well as on longer time scales. The
overall count rate in 2003 (C2) is nearly three times as high as in
2000 (C1). In the two consecutive observations X2 and X3, the X3 count rate
is nearly twice as high as the X2 count rate. The X3 luminosity curve 
(Fig.~\ref{lcslumi}) starts with an increase followed by a slight diminution in 
emission towards the end of the observation. The high-luminosity points at the 
beginning of the X2 luminosity curve remain unexplained.
The X-ray spectra (Fig.~\ref{xrayspectra}) show signs of high absorption 
$(N_H \approx 4 \times 10^{22}$~cm$^{-2}$), the temperature of the emitting
plasma is at around $T \approx 3 \times 10^7$~K. There appears to be no 
significant temporal evolution in these fit parameters (Fig.~\ref{xrayfitpars_v} and Table~\ref{fitrestab}).
\citet{nis05} confirm earlier observations by \citet{che93}, finding that IRS~5 is
a binary object with 112~AU separation (scaled to $d=150$~pc). Thus, the attribution of
X-rays to any of the components, or both is difficult. For IRS~5a, they find that
accretion accounts for below 20\% of the object's bolometric luminosity.
While this suggests a deeply embedded object in a later evolutionary stage, 
\citet{nis05}, using standard evolutionary tracks, conclude that IRS~5a has about
the same age as IRS~2 ($\approx 5 \times 10^5 - 10^6$~yrs).
Already \citet{fem99} note that the radio properties of this source
rather resemble a weak-line T Tauri star. Due to the compact grouping
of protostars surrounding it, it remains unclear whether IRS~5
powers an outflow (e.g. \citealp{wan04}). The weak neighbouring radio
source (no. 5 in Tables~\ref{sourceidlist} and \ref{sourceralist}) has 
a very faint 2MASS counterpart within the nebulosity surrounding IRS~5.

\vspace*{3mm}
\noindent
IRS 1 was not found to be variable at radio frequencies of 4.9 GHz  \citep{sut96}. In our data, this source does not
show significant radio variability either (Fig.~\ref{vlacurves} and Table~\ref{sourceralist}). The observed flux is compatible with
the value measured by \citet{fcw98}. There are, however, clear signs
of X-ray variability in this relatively bright X-ray
source -- within the datasets as well as on longer time scales. 
The C2 count rate is 35\% greater than the C1 count rate, and the 
XMM-\textsl{Newton} count rates also suggest variability even though the
X2 and X3 count rates are affected by a chip gap in the source region, which
causes a flux loss af about 20\%. The X-ray luminosity curves (Fig.~\ref{lcslumi})
show only slight variability within the single observations.
The X-ray spectra (Fig.~\ref{xrayspectra}) 
can be modeled by highly absorbed $(N_H \approx 3.5 \times 10^{22}$~cm$^{-2})$ 
emission of hot plasma $(T \approx 5 \times 10^7$~K). There is a significant
difference in the absorbing column densities derived from the X1 and C2 observations.
No such signs of differences in the fitted plasma temperatures are seen 
(Fig.~\ref{xrayfitpars_v} and Table~\ref{fitrestab}).
According to \citet{wan04}, IRS~1 could be the powering source of four 
Herbig-Haro objects. \citet{nis05} find that 80\% of this object's 
bolometric luminosity is due to accretion (see also \citealp{nis04} who find
evidence for an accelerating wind in the vicinity of IRS~1).

\vspace*{3mm}
\noindent
IRS~9 is detected as a weak X-ray source throughout all
datasets studied here, shows an X-ray flare in the
C1 \textsl{Chandra} data (not shown), however, was not
detected at radio wavelengths ($S<0.035$~mJy). This is consistent
with previous radio observations.

\subsubsection{Radio Source 5 and IRS~6: A candidate brown dwarf and a T Tauri star}

Radio Source 5, initially observed by \citet{bro87}, has not
been detected at X-ray wavelengths. In our data, its radio intensity is
variable on timescales of days to months (Fig.~\ref{vlacurves} and Table~\ref{sourceralist}).
\citet{fcw98} report a radio flux on a similar scale. Variability of
this source at radio frequencies (4.9 GHz) was also reported by \citet{sut96} 
who argue against an extragalactic origin of the emission. Given its faint
infrared counterpart, this source could be a brown dwarf \citep{wil97,fcw98}. If confirmed, the radio data presented here together with the data analyzed by \citet{sut96} would constitute an interesting dataset of long-term variability of brown dwarf radio emission.

\vspace*{3mm}
\noindent
IRS 6, detected as a weak X-ray source by \textsl{Chandra} as well as
XMM-\textsl{Newton}, was only found in the combined AM596 $uv$-data: Its 
emission is very weak at 0.15~mJy. No radio emission from this source
has been reported before. While this source is too weak for an
analysis of short-timescale X-ray variability, the emission is
somewhat weaker in the 2003 (C2) data when compared to the 2000 (C1)
level. The X2 and X3 count rates for IRS~6 are affected by a chip 
gap in the source region.
\citet{wan04}, looking for optical
outflows, argue that IRS~6 could be the driving source of four
Herbig-Haro objects. \citet{nis05} resolve this
source into a binary object with a separation of only 78 AU, making
the attribution of X-rays difficult. However, they find no sign of
ongoing accretion in the near-infrared. Thus, IRS~6 appears to
contain at least one T Tauri star. 

\subsubsection{The IRS 7 complex with its two deeply embedded protostars}

Two main sources can be detected in the IRS~7 region, at both radio
and X-ray wavelengths, namely IRS~7E and IRS~7W. Another major radio source is located five arcseconds north of IRS~7W (Radio Source 9, as defined by \citealp{bro87}). Three additional weak radio sources accompany IRS~7E/W.

There has been confusion about possible protostellar sources
detected at different wavelengths around IRS~7, although some
clarification was brought by \citet{nut05}, presenting
submillimeter data where the single emission peak observed before at
millimeter wavelengths \citep{hen94, chi03, gro04} is resolved into
three subsources. A molecular outflow and a disk were detected 
in the IRS~7 region and analyzed by \citet{and97} and \citet{gro04},
although the latter showed that it is not yet possible to attribute these to any of the sources seen at higher angular resolution.
IRS~7 is also the driving source of Herbig-Haro objects \citep{wan04}.
\citet{har01} find evidence for a jet emanating from IRS~7E at 
$\lambda=6$~cm, using the ATCA radio interferometer.
\citet{cho04} analyze the IRS~7 region at $\lambda=6.9$~mm with the VLA. 
Interestingly, they only marginally detect IRS~7E while there is elongated emission around IRS~7W and Radio Source 9, although the connection remains unclear. The emerging picture is that of IRS~7W being an infrared-detected, deeply embedded protostar, probably a class~I or II source, while IRS~7E could be a class~0 source although the source was detected at 4.8~$\mu$m by \citet{pon03}. This would be one of the first class~0 sources
detected in X-rays.

IRS~7E is quite variable on all timescales covered in the 1998 VLA data (Fig.~\ref{vlacurves} and Table~\ref{sourceralist}), at a flux level mostly below the value reported by \citet{fcw98}. The weak radio source detected in the eastern vicinity of IRS~7E (source 15) might be a sign for extended radio emission. The radio sources detected around IRS~7E/W are shown in an inset of Fig.~\ref{vlainset}.

\begin{figure}
% \begin{center}
  \includegraphics*[width=9cm]{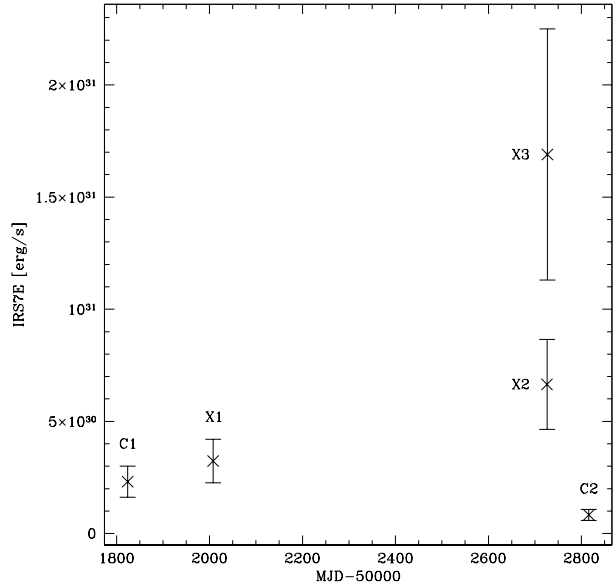}
  \caption{IRS~7E X-ray luminosity. Countrates for the five X-ray datasets 
  were converted into luminosities without background subtraction, assuming
  a single-temperature spectrum of $kT=3$~keV and $N_H=2.8 \times 
  10^{23}$~cm$^{-2}$ \citep{ham05}. The errorbars denote errors of 30\%. 
  Note that the source is only marginally detected in C1 and C2.}
  \label{irs7elumi}
% \end{center}
\end{figure}

IRS~7E is the most variable X-ray source among the objects studied here. While IRS~7E was only marginally detected
in the C1, X1, and C2 observations, the 2003 (X2, X3) datasets
show IRS~7E as a strong and very hard X-ray source. This flaring has been 
studied by \citet{ham05} who argue that this might be an X-ray detected
class~0 source. They do not find any near-infrared counterpart down to a $K$-band magnitude of 19. However, \citet{pon03} detect both sources, IRS~7E/W, at 4.8~$\mu$m using VLT-ISAAC, arguing that IRS~7E remained undetected at 10~$\mu$m \citep{wil97} due to silicate absorption.
It is tempting to attribute also the enormous flare observed
by \citet{koy96} to this source, although that remains unclear due to the limited angular resolution of their ASCA data.
Since spectra of IRS~7E could only be taken from the XMM-\textsl{Newton} data, an estimated luminosity curve for this source was produced by assuming the same spectral characteristics, taken from \citet{ham05}, of $kT=3$~keV (corresponding to a temperature of $\approx 35$~MK, rather a lower limit) and $N_H=2.8 \times 10^{23}$~cm$^{-2}$ for all five datasets, then determining the fluxes and luminosities following the procedure outlined above. To minimize contamination, \citet{ham05} excluded IRS~7W from the IRS~7E source region in their analysis. The plot shows the enormous increase in luminosity (Fig.~\ref{irs7elumi}). In the two \textsl{Chandra} observations, only 19 and 13 -- however highly energetic -- photons were recorded from IRS~7E respectively. Given the high temperature of the plasma emission as well as the quick variability, especially when comparing the subsequent X2 and X3 datasets, it appears unlikely that the emission is due to an accretion shock. These characteristics rather point towards circumstellar magnetic activity.

IRS~7W is the brightest radio source in the sample considered here, mostly emitting at around 4.2~mJy, although in two epochs the source is detected at $>5$~mJy (Fig.~\ref{vlacurves} and Table~\ref{sourceralist}). This source shows only moderate radio variability within the 1998 monitoring, especially on timescales of days.
\citet{fcw98} report an 8.4 GHz flux density of 1.58~mJy, considerably higher than all fluxes measured in 1998 which are $<1.1$~mJy. Close to Radio Source 9, there are two weak radio sources (sources 10 and 13) which do not have counterparts at other wavelengths. 

In the C1/C2 data, IRS~7W is detected as a weak X-ray source emitting energetic photons, similar to IRS~7E. Due to contamination caused by the flaring nearby source IRS~7E, a meaningful spectrum could only be extracted from the X2 dataset, leading to $kT=4.7$~keV ($\approx 55$~MK) and $N_H=3.4 \times 10^{23}$~cm$^{-2}$\citep{ham05}.

\subsubsection{The Herbig Ae star R CrA}

R CrA is one of the optically brightest ($V \sim 11.5$~mag)
objects in the CrA cloud and it illuminates the reflection nebulosity
NGC~6729. The stellar parameters are somewhat uncertain, especially because
the optical spectrum seems to be variable and spectral types ranging from
B8 to F7 have been reported (e.g.~\citealp{bib92}). Here we adopt the
values listed in \citet{lor99}, i.e.~spectral type A5e, 
$L_{\rm bol} = 132\,{\rm L}_\odot$, and $A_V = 1.9$~mag.
Comparison of these parameters to the pre-main sequence models
of \citet{pal99} suggest
a mass of $\sim 3.5\,{\rm M}_\odot$ and an age of slightly more than
1~Myr for R~CrA.

The spectral energy distribution of R~CrA rises very steeply between 
$0.6\,\mu$m and $3\,\mu$m (see, e.g.~\citealp{ack04}), 
demonstrating the
presence of large amounts of hot circumstellar material, presumably 
in the form of an optically thick circumstellar disk. This very large 
infrared excess suggests that R~CrA is in a very early evolutionary state
and thus we discuss it here together with the lower-mass protostars.

In the VLA data (Fig.~\ref{vlacurves} and Table~\ref{sourceralist}), R~CrA was detected as a weak and rather constant
radio source with an average flux of 0.23~mJy, the same flux level as measured by
\citet{fcw98}. However, R~CrA remained undetected at radio wavelengths 
several times before: in September 1985 (VLA, 6cm, 3$\sigma$=0.17~mJy, 
\citealp{bro87}), in February 1990 (VLA, 3.6cm, 3$\sigma$=0.10~mJy, 
\citealp{ski93b}) as well as with the VLA and the AT in 1985-87 and 
1992, respectively \citep{sut96}. Even though R~CrA is quite close to 
the respective 5$\sigma$ limits, this is suggestive of long-term 
variability.

\begin{figure*}
%\begin{center}
\centering
\begin{minipage}{4cm}
      \centerline{\hbox{
      \includegraphics*[width=3.9cm]{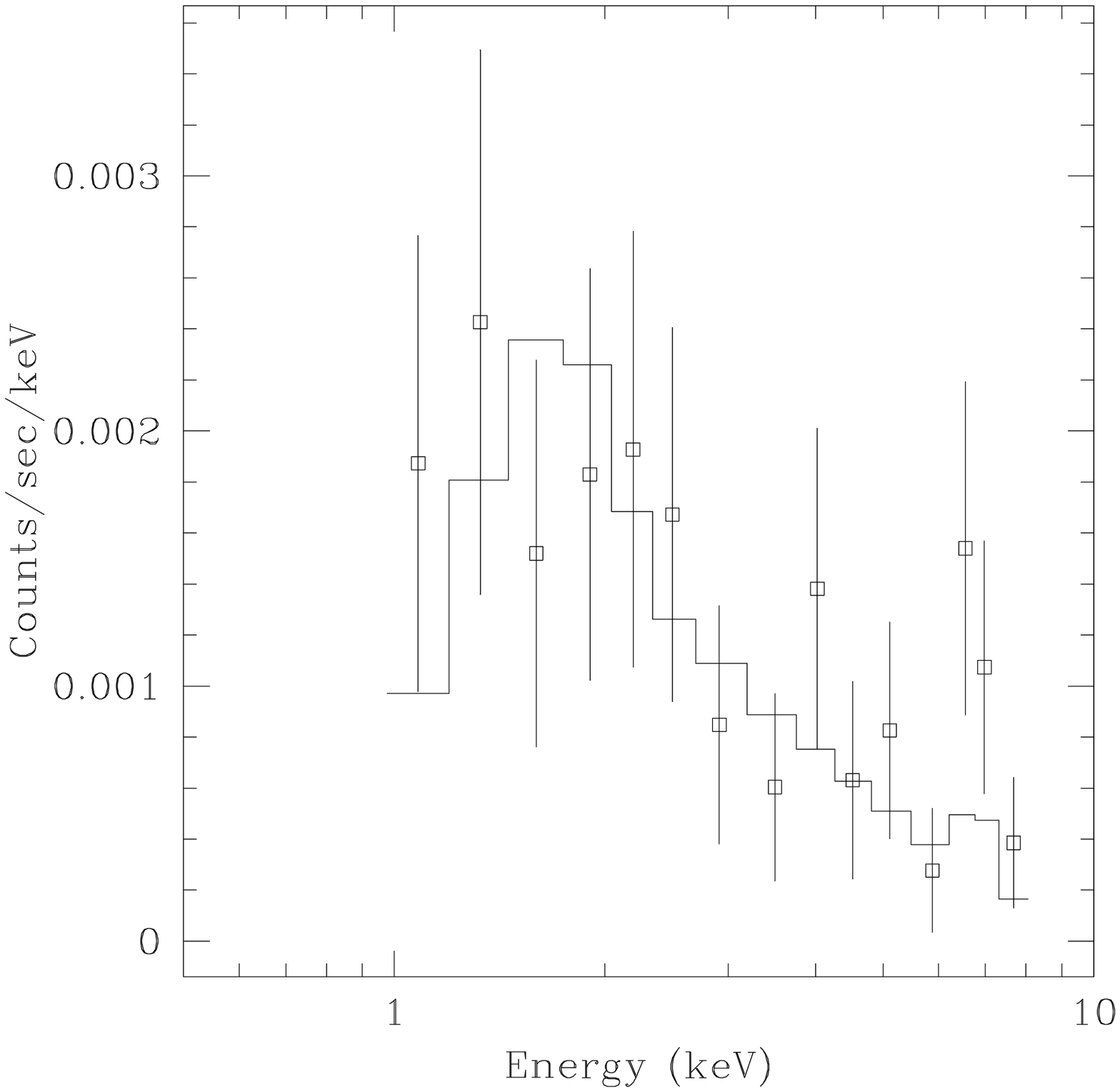}
      }}
      \centerline{X1}
\end{minipage}\    \
\begin{minipage}{4cm}
      \centerline{\hbox{
      \includegraphics*[width=3.9cm]{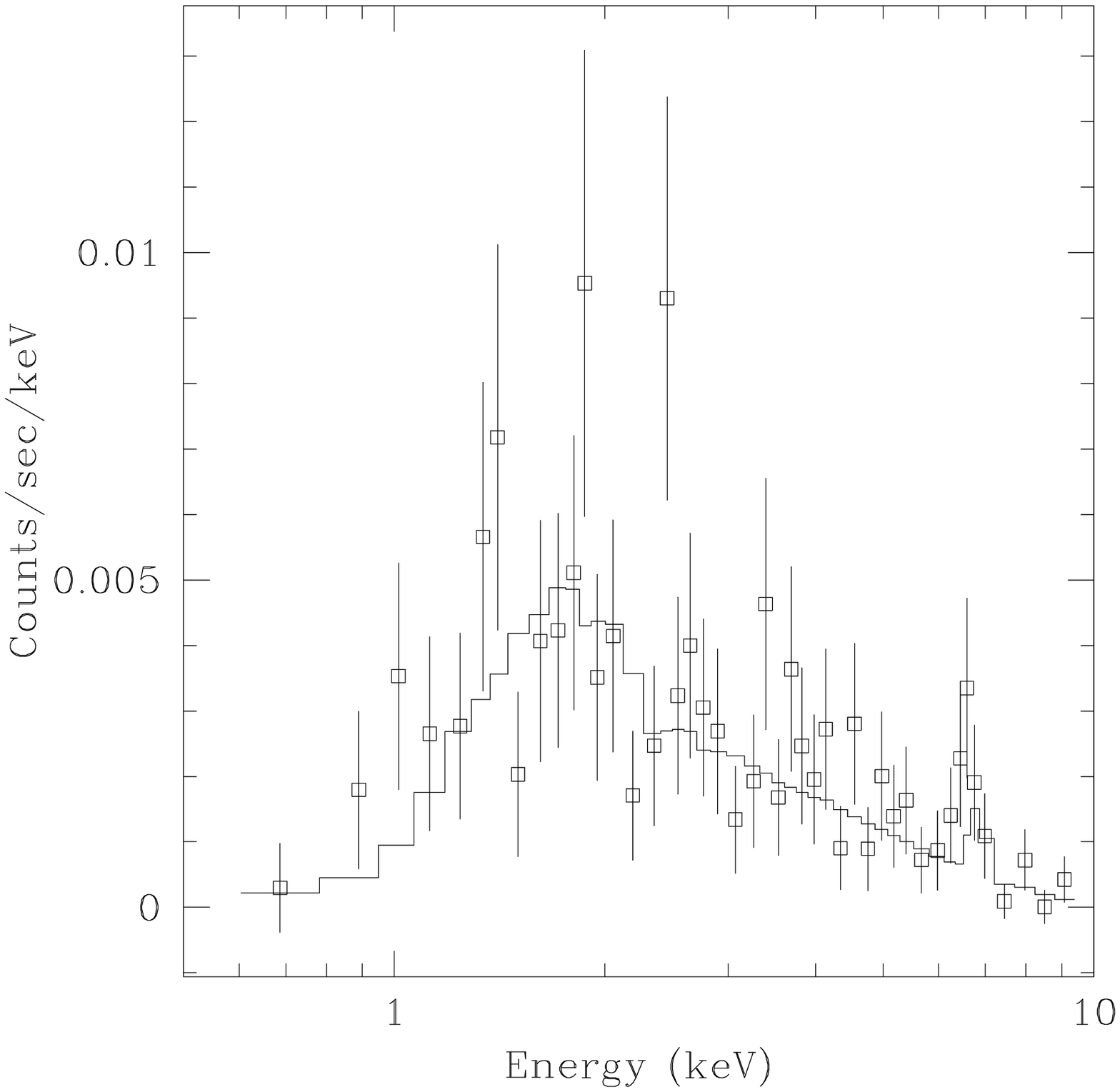}
      }}
      \centerline{X2}
\end{minipage}
\begin{minipage}{4cm}
      \centerline{\hbox{
      \includegraphics*[width=3.9cm]{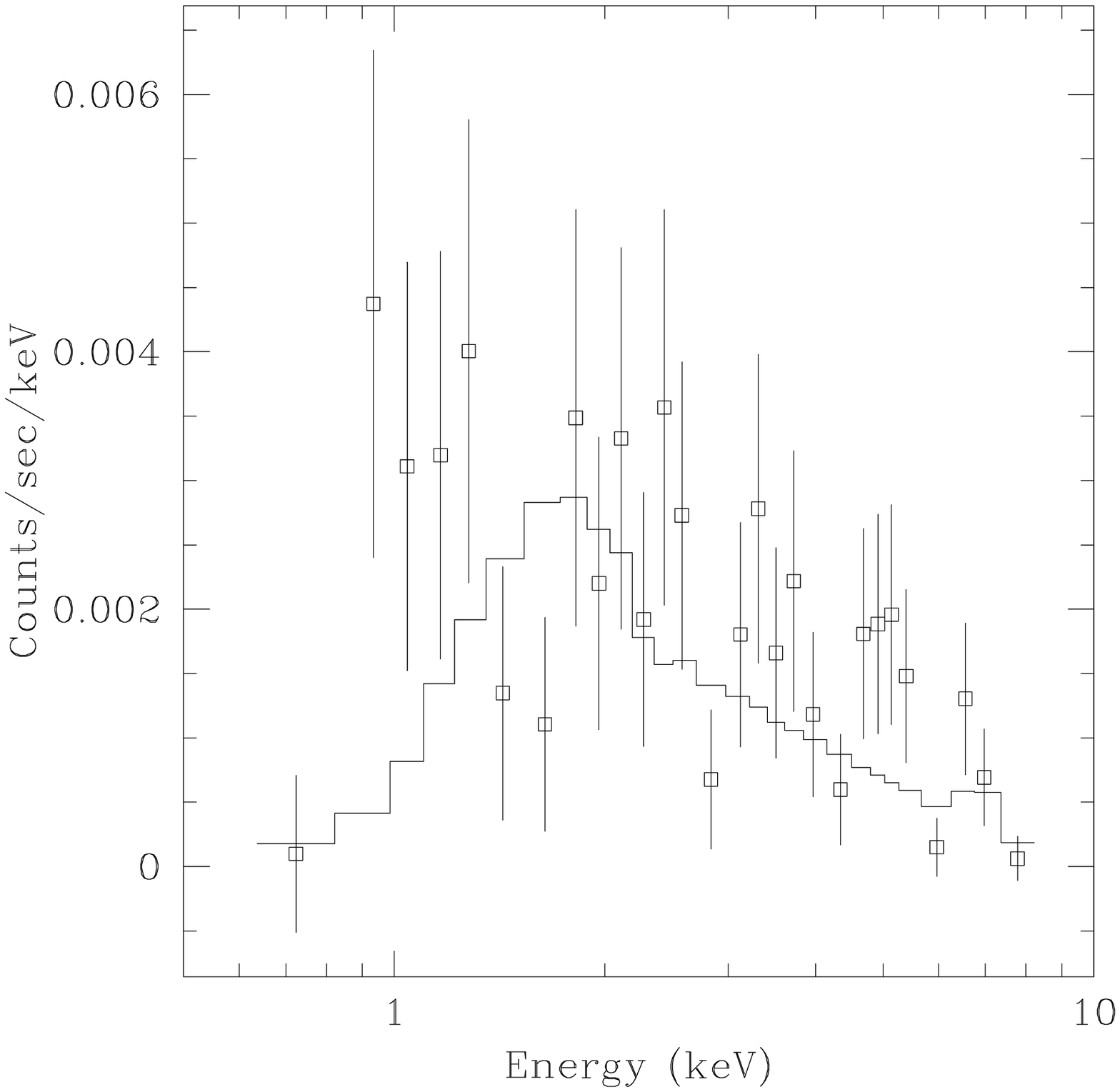}
      }}
      \centerline{X3}
\end{minipage}
%\end{center}
\caption{R CrA spectra, taken from an ellipse region avoiding the neighbouring
source IRS~9. The results of fitting an absorbed APEC model are
shown. In the X3 fit, the column density was frozen to the values from X1 and X2.}
\label{rcraspec}
\end{figure*}

\begin{figure*}
% \begin{center}
\centering
  \includegraphics*[width=16cm, bb=10 480 593 710]{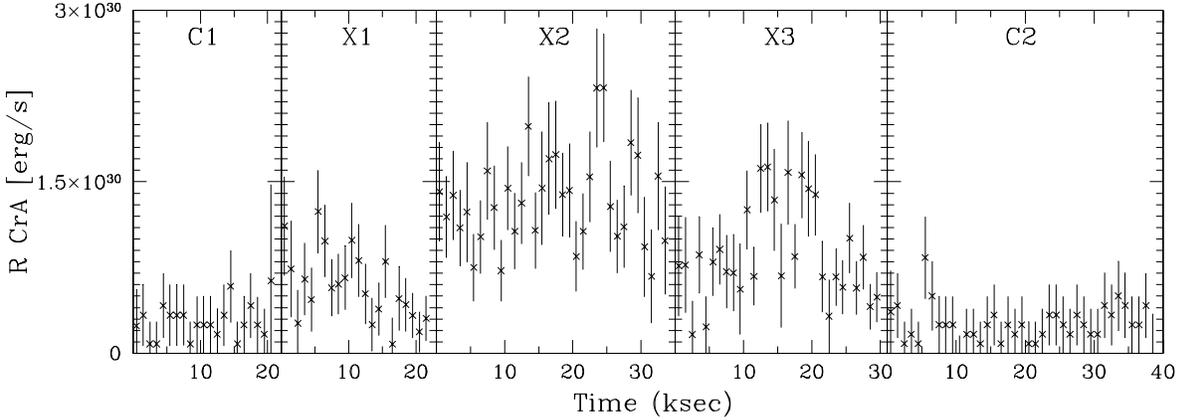}
  \caption{R~CrA X-ray luminosity curves from the same elliptical region also used
  for the spectra. Luminosity calculated assuming spectral properties of the best
  spectrum (X2). Data is kilosecond-binned and background-subtracted, datasets are
  in temporal order as labeled.}
  \label{rcralumi}
% \end{center}
\end{figure*}

X-ray observations of the \textsl{Coronet} region with EINSTEIN and ROSAT
did not detect X-ray emission from R~CrA \citep{zip94,dam94,nep97,wal97}.
\citet{koy96} detected hard X-ray emission near the location
of R~CrA with ASCA. However, due to the poor spatial resolution of the ASCA data
(FWHM $\sim 2'$) the contributions of the individual young stellar objects
R~CrA, IRS~7, and IRS~9  could not be resolved. Nevertheless, since
the intensity peak in the ASCA image was located close to the position
of R~CrA, it appears likely that R~CrA produced a significant fraction of the
observed hard X-ray emission.
During this ASCA observation a strong flare was seen.
The X-ray spectrum was found to be very hard
and could be fitted with a $kT \sim 6-7$~keV plasma model
and an absorbing hydrogen column density of $N_{\rm H} \sim 4\times 10^{22}\,
{\rm cm}^{-2}$. 
During the flare, the spectrum showed a remarkable broadened or double emission
line structure between 6.2 and 6.8~keV. However, it remained unclear
which of the sources caused the flare.

X-ray emission from R~CrA is clearly detected in all the \textsl{Chandra} and
XMM-\textsl{Newton} data sets used in this paper. The detection in the 
\textsl{Chandra} observation C1 was already noted by \citet{ski04}.
During both \textsl{Chandra} observations R~CrA is only seen as a rather weak
source, yielding no more than $\sim 100$ counts, too few for detailed
spectral analysis. During the first XMM-\textsl{Newton} observation the observed
count rate for R~CrA is well consistent with that in the \textsl{Chandra} 
observations, but during the XMM-\textsl{Newton} observations X2 and X3 the 
count rates were
about a factor of two larger (see lightcurve in Fig.~\ref{rcralumi}).
The lightcurves show no indications of a flare, it appears rather 
as if the source has ``switched'' to a higher activity level during X2 and X3.
A sudden drop of the count rate (by about a factor of two) 
is seen in the second half of observation X3. A possible interpretation
for the general shape of the 
lightcurve would be the temporal presence of a dominant active region 
during X2 and X3. The sharp drop in X3 may then be explained by the
occultation of such an active region when the stellar rotation moves it
behind the stellar disk.

The X-ray spectra of R~CrA extracted from observations X1, X2, and X3
show a very strong high-energy tail and a pronounced 6.7~keV emission
line, providing clear evidence for very hot plasma.
The spectral fits yield a plasma temperature of about 100~MK, the highest
value of all sources in the \textsl{Coronet} region. The
hydrogen column densities derived from the spectral fits 
are $\sim 1.3\times 10^{22}\,{\rm cm}^{-2}$ and correspond to an 
optical extinction of $A_{V,X} \sim 6 - 8$~mag \citep{ryt96,vuo03}.
This is higher than the extinction estimate derived from the optical properties
($A_V \sim 2$~mag), but given the uncertainty of the optical parameters
of R~CrA, both values may well be in agreement.

The X-ray luminosity of R~CrA derived from the spectral fits ranges
from $2.5 \times 10^{29}$~erg/sec in the C2 observation to $1.3 \times 10^{30}$~erg/sec
in the X2 observation.
The fractional X-ray luminosity is thus 
$\log\left(L_{\rm X}/L_{\rm bol}\right) \sim -6.3 \dots -5.6$.

The observed X-ray emission of R~CrA is particularily interesting since, 
according to our current understanding of stellar X-ray emission mechanisms, 
one would not expect to see any  X-ray emission emission from this star.
Being an intermediate-mass A-type star, R~CrA should neither possess a
magnetically driven corona, which is the source of the X-ray emission 
from late type (F to M) stars, nor can it have
a strong radiation driven stellar wind, in which shocks
cause X-ray emission observed in the more massive O- and early B-type stars 
(see e.g.~\citealp{pal89,fam03}).
The theoretical expectation that late B and A stars should not be 
instrinsic X-ray emitters is well confirmed in the case of main sequence stars
(e.g.~\citealp{sch93,cas94}).
However, it is still unclear whether the Herbig Ae/Be stars, i.e.~the class of
very young intermediate-mass pre-main sequence stars, to which R~CrA belongs,
also fit to this scenario. In EINSTEIN, ROSAT, ASCA, and \textsl{Chandra}
 X-ray observations of samples of Herbig Ae/Be stars 
\citep{dam94,zip94,prz96,ski04,ham05} 
some 30 - 50~\% of the observed stars were detected as X-ray sources. 
Several possible explanations have been proposed for these findings
(see \citealp{cai94, prz96, ski04}),
but no final explanation has been found. There exist some ideas how 
an AeBe star may produce X-ray emission (e.g.~the scenario of
a nonsolar dynamo powered by rotational shear in very young intermediate 
mass stars; see \citealp{top95}),
but no really convincing evidence has yet been found for this.
A possible explanation for the observed X-ray emission 
may be that the X-rays originate
not from the AeBe stars themselves, but from close, unresolved,
late type companions. 

For our case of R~CrA, we note that
\citet{tak03} presented spectro-astrometric observations
and found some evidence for the presence of a
companion at a separation of about $\sim 0.070``-0.10``$, i.e. only 10--15~AU.
If such a companion actually exists, it is probably a lower-mass
young stellar object, i.e.~a T Tauri star or a low-mass protostar,
which could easily explain the observed X-ray emission.

What remains quite remarkable, however, is the extremely high plasma temperature
of $\sim 100$~MK inferred from the X-ray spectra. Typical plasma temperatures
in young stellar objects range from $\sim 10$~MK up to $\sim 50$~MK, 
reaching higher values usually only during large X-ray flares\footnote{
In the most comprehensive data set on the X-ray emission from young stars,
the Chandra Orion Ultradeep Project, only 2\% of the optically visible T Tauri stars in the Orion Nebula Cluster showed plasma temperatures exceeding $100$~MK 
(see \citealp{get05,pre05}).}.
From other X-ray observations of star forming regions it seems that
class~I protostars show systematically higher plasma temperatures 
than the more
evolved T~Tauri stars (see, e.g., \citealp{ikt01} for the
case of $\rho$~Oph). This may be an indication that the companion of
R~CrA is a very young object (i.e.~more likely a class~I protostar than
a T Tauri star). In this case, some (perhaps the largest) fraction of the
infrared excess emission may come from the companion and not from
the Ae star R~CrA itself.

\section{Conclusions and Outlook}

We presented the results of radio monitoring of protostars in the
\textsl{Coronet} cluster together with archival X-ray data of these
sources covering 154~ksec. The main results are:

\begin{enumerate}

\item Observing with the VLA in nine epochs spread over several 
months in 1998, we detect centimetric radio emission from 13 sources in the
region around R~CrA. Nine sources in this region are detected in X-rays, and
eight sources are detected at both radio and X-ray wavelengths.

\item IRS~5, a class I protostar with nonthermal radio emission, shows an interesting
increase and decrease in its radio emission on the order of days, accompanied
by changes in its polarization properties. It is the most variable 
radio source in the 1998 VLA data. Among the three protostars bright 
enough for a detailed X-ray study, IRS~5 is also the most variable source in
X-rays.

\item Not counting IRS~7W due to the unclear classification, we
detect three out of four covered class~I sources in radio
emission, while we detect all four in X-rays. With IRS~7E, one candidate class~0 source is detected in X-rays as well as in radio emission.

\item The Herbig Ae star R~CrA was detected at both radio and X-ray
wavelengths. The X-ray spectrum of this source surprisingly contains
emission from extremely hot plasma ($\approx 9 \times 10^7$~K). We suspect
that a lower-mass companion is responsible for this emission.

\item The spectra of IRS~1, 2, and 5 all can be explained by the
absorbed emission of plasma having temperatures of several $10^7$~K
while the high absorbing column densities (several
$10^{22}$~cm$^{-2}$) are all at about half the values determined from
near-infrared colours. While this has been observed before towards similar
sources, the effect remains unexplained. We discuss possible explanations.

\item While the X-ray lightcurves of the the class I protostars IRS~1, 2,
and 5 all show varying degrees of variability, no flare events were observed.
With no class II sources to compare with, it is impossible here to corroborate
the conclusion of \citet{ikt01} that class I sources are intrinsically more X-ray--variable than later evolutionary stages, although the three class I sources studied here are not extremely variable.

\item The X-ray spectra are spread over about 2.5 years and also allow for
an analysis of variability in their main fit parameters, absorbing column
density and plasma temperature. While mostly, no significant variability can
be seen, a significant change is observed in the absorbing column density towards IRS~2: It appears to have decreased continuously to a significantly lower level, from $2.4 \times 10^{22}$~cm$^{-2}$ to $1.5 \times 10^{22}$~cm$^{-2}$, possibly due to a clumpy circumstellar medium towards this source.

\end{enumerate}

Studying the \textsl{Coronet} cluster at radio and X-ray wavelengths has
provided a number of interesting insights. In order to learn more about
high-energy processes in these young stellar objects, we plan to carry 
out simultaneous X-ray, radio, and near-infrared observations of the \textsl{Coronet} cluster in 2005 with \textsl{Chandra}, the VLA, and at ESO.

\begin{acknowledgements} 
The National Radio Astronomy Observatory is a facility of the National 
Science Foundation operated under cooperative agreement by Associated 
Universities, Inc. This study is partly based on observations obtained with 
XMM-\textsl{Newton}, an ESA science mission with instruments and 
contributions directly funded by ESA Member States and NASA. 
\end{acknowledgements}

%\begin{thebibliography}{}

\bibliographystyle{aa} % style aa.bst
\bibliography{rcra} % your references Yourfile.bib

%\end{thebibliography}

\end{document}